\documentclass{aa}  
\usepackage[T1]{fontenc}
\usepackage{txfonts} 
\usepackage{graphicx}
\usepackage{xspace}
\usepackage{booktabs}
\usepackage{tabularx,multirow,booktabs,blindtext}
\usepackage{algorithm}
\usepackage{algpseudocode}
\usepackage{amsmath}
\usepackage{ulem}
\usepackage{xcolor}
\usepackage{listings}

\providecommand{\keywords}[1]{\textbf{\textit{Index terms---}} #1}

\providecommand{\keywords}[1]{\textbf{\textit{Index terms---}} #1}

\newcommand{\hMpc}{\ensuremath{h\;\text{Mpc}^{-1}}\xspace}
\newcommand{\kpch}{\ensuremath{h^{-1}\,\text{kpc}}\xspace}
\newcommand{\Mpch}{\ensuremath{h^{-1}\,\text{Mpc}}\xspace}
\newcommand{\Lya}{Lyman-$\alpha$\xspace}
\newcommand{\sigHI}{\sigma_{\mathrm{HI}}}
\newcommand{\LCDM}{$\Lambda$CDM\xspace}
\newcommand{\mvec}[1]{\mathbf{#1}}
\newcommand{\Msun}{\ensuremath{\mathrm{M}_\odot}\xspace}
\newcommand{\LyNet}{\textit{LyAl-Net}\xspace}
\newcommand{\HnoAGN}{\textsc{Horizon-noAGN}\xspace}
\newcommand{\HAGN}{\textsc{Horizon-AGN}\xspace}
\newcommand{\HDM}{\textsc{Horizon-DM}\xspace}

\newcommand{\nHI}{\ensuremath{{n_\text{HI}}}\xspace}
\newcommand{\TempHI}{\ensuremath{{T_\text{HI}}}\xspace}
\newcommand{\ramses}{{\sc Ramses}\xspace}
\begin{document}

   \title{LyAl-Net: A high-efficiency Lyman-$\alpha$ forest simulation with a neural network}
   \titlerunning{\textit{LyAl-Net}}

   \subtitle{}

   \author{Chotipan Boonkongkird
          \inst{1}
          \thanks{chotipan.boonkongkird@iap.fr}
          \and
          Guilhem Lavaux\inst{1} 
          \and
          Sebastien Peirani\inst{1,2}
          \and
          Yohan Dubois\inst{1}
          \and
          Natalia Porqueres\inst{3}
          \and
          Eleni Tsaprazi \inst{4} 
          }

   \institute{Sorbonne Université, CNRS, UMR 7095, Institut d’Astrophysique de Paris, 98 bis bd Arago, 75014 Paris, France 
        \and Université Côte d’Azur, Observatoire de la Côte d’Azur, CNRS, Laboratoire Lagrange, Bd de l’Observatoire,
CS 34229, 06304 Nice Cedex 4, France
        \and Department of Physics, University of Oxford, Denys Wilkinson Building, Keble road, Oxford OX1 3RH, United Kingdom
        \and The Oskar Klein Centre, Department of Physics, Stockholm University, Albanova University Center, SE 106 91 Stockholm, Sweden     }

   \date{Received 30 March 2023}

% \abstract{}{}{}{}{} 
% 5 {} token are mandatory
 
  \abstract
  % context heading (optional) 
   {The inference of cosmological quantities requires accurate and large hydrodynamical cosmological simulations. Unfortunately, their computational time can take millions of CPU hours for a modest coverage in cosmological scales ($\approx (100 \ensuremath{h^{-1}}\,\text{Mpc})^3)$). The possibility to generate large quantities of mock Lyman-$\alpha$ observations opens up the possibility of much better control on covariance matrices estimate for cosmological parameters inference, and on the impact of systematics due to baryonic effects.}%
% Aims
   {We present a machine learning approach to emulate the hydrodynamical simulation of intergalactic medium physics for the Lyman-$\alpha$ forest called \textit{LyAl-Net}. The main goal of this work is to provide highly efficient and cheap simulations retaining interpretation abilities about the gas field level, and as a tool for other cosmological exploration. }%
%Methods
   {We use a neural network based on the U-net architecture, a variant of convolutional neural networks, to predict the neutral hydrogen physical properties, density, and temperature. We train the \textit{LyAl-Net} model with the \textsc{Horizon-noAGN} simulation, though using only 9\% of the volume. We also explore the resilience of the model through tests of a transfer learning framework using cosmological simulations containing different baryonic feedback.   We test our results by analysing one and two-point statistics of emulated fields in different scenarios, as well as their stochastic properties.
   }%
% Results
   {The ensemble average of the emulated Lyman-$\alpha$ forest absorption as a function of redshift lies within 2.5\% of one derived from the full hydrodynamical simulation. The computation of individual fields from the dark matter density agrees well with regular physical regimes of cosmological fields. The results tested on IllustrisTNG100 showed a drastic improvement in the Lyman-$\alpha$ forest flux without arbitrary rescaling. }
% Conclusions
   {Such emulators could be critical for the exploitation of upcoming surveys like WEAVE-QSO. The transfer learning technique is showing promises to alleviate sensitivity to the learned physical model. This technique could have a decisive impact on the results derived from present experiments, such as QSO surveys derived from SDSS3, which has a resolution power $R=1500$, and SDSS4 ($R=2000$) data, which are the resolution ranges available with our emulator.}

\maketitle

\section{Introduction}
\label{sec:intro}
The \Lya forest is a collection of sharp absorption features, generally observed from distant Quasi-Stellar object (QSO) spectra, first observed by \citet{Lynds1971}. This phenomenon arises from the photons emitted from background sources, with frequencies higher or equal to \Lya frequency, undergoing redshift due to the Hubble expansion and absorbed by the intervening intergalactic medium. Since the absorptions occur at different distances from an observer in a line-of-sight, they are also redshifted. This mechanism creates a forest-like ensemble of features on the observed QSO broadband \Lya emission. 
While the process only happens with neutral hydrogen (HI) atoms, which represent a small fraction of the intergalactic medium (IGM), one can use the properties of the \Lya forest to infer the spatial structure of the baryonic matter in the universe. For the concordance $\Lambda \mathrm{CDM}$ model, the majority of matter in the universe is under the form of dark matter (DM) \citep{Jarosik_2011,Planck2020}. 

Thus the baryonic spatial distribution is mainly dictated by the gravitational potential imprinted by the dark matter density. Therefore, \Lya forest observations are good tracers of the total matter in the context of this cosmological model \citep{Petitjean_1995,Hui1997,Croft_1998}.  Moreover, it is crucially important for cosmology, as at the relevant redshifts probed by \Lya, we do not have as many easily observable galaxies as at low redshift ($z \lesssim 1$). The \Lya forest is thus a complementary probe of the matter clustering at $z\simeq 2-3$, also known as the cosmic noon era.  

The \Lya forest can provide much insight into the cosmological model. Past work on the subject includes, for example, the recovering of the properties of the intergalactic medium using the Bayesian inversion method \citep{Pichon2001}, the constraining of the neutrino masses \citep{Palanque_Delabrouille_2015,Yche_2017}, the inference of the three-dimensional matter distribution from 1D \Lya absorption using Bayesian Forward modelling \citep{Porqueres_2019,Horowitz2019,Porqueres_2020,Kraljic2022}, the constraints on the thermal history of IGM \citep{Villasenor_2022}, and the measurement of the  Baryonic Acoustic Oscillation (BAO), which provides a tight constraint on the expansion history of the universe \citep{Font_Ribera_2014}. %and other cosmological parameters
%DATA observation part & applications 
Several large cosmological surveys include nowadays a QSO survey with spectrum measurement \citep[BOSS, eBOSS, DESI][]{Lee_2013,du_Mas_des_Bourboux_2020,DESI}.

To analyse the \Lya forest observations, we require hydrodynamical simulations of the IGM, which has to be computed alongside a dark matter $N$-body simulation. Though the IGM is coupled on large scales with the gravitational potential induced by dark matter overdensity, on small scales, it is dominated by baryonic shocks, cooling, and feedback, for which we do not have analytical solutions. Therefore, the IGM state has to be solved numerically. With upcoming cosmological surveys such as WEAVE-QSO \citep{WEAVE-QSO}, the resolution and volume become higher and larger. They consequentially increase the required computational time drastically. Therefore using full hydrodynamical simulation to match the same volume is becoming intractable.
Alternatively, $N$-body simulations of pure dark matter are much faster than hydrodynamical simulations.
Thus, one could avoid solving for the hydrodynamic altogether and trying to correct phenomenologically the effect at small scales ($\sim 100\kpch$, the Jeans' scale where \Lya happens). 

The fluctuating Gunn-Peterson approximation \citep[FGPA,][]{Gunn1965,Weinberg1997} is a main framework to estimate a mock \Lya from dark matter overdensity. However, this approximation does come with a few limitations. For example, it fails to capture the baryonic feedbacks at small scales, the power-law breaks down in the regions with higher density and strongly heated gas \citep{Luki__2014,Kooistra_2022}, and the power law starts to decouple where $z \gtrapprox 3$. In most cases, the dark matter overdensity is smoothed before being used for emulation. For example,  the baryonic pressure is emulated by convolving the matter density with a Gaussian kernel \citep{Hui1997}. Other work \citep{Sorini_2016} introduced a numerical simulation called Iteratively Matched Statistics (IMS) with an effort to improve the accuracy higher than Gaussian smoothing. Finally, the LyMAS-2 technique \citep{Peirani_2014,Peirani_2022} relies on a conditional probability to map the flux from a smoothed dark matter field and introduces a Wiener filter to represent the coherence along line-of-sight better. 

In recent years, the astrophysical community growingly adopted machine learning models because of their ability to provide a universal fitting function and its prediction speed through accelerators such as Graphics Processing Units (GPUs), which allows us to reach both a highly accurate and fast model, provided training data exist.
Inspired by the work of \cite{Peirani_2014}, we propose to swap the conditional PDF with a neural network. In this work, we will test a \textit{U-Net} architecture, focusing on emulating the most essential fields to describe the IGM for \Lya. We introduce such an emulator, called \LyNet, trained to map a dark-matter density field from an $N$-body simulation to IGM fields  (neutral hydrogen density and temperature) derived from a sibling hydrodynamical simulation. In this work, the two simulations are Horizon-DM and \HnoAGN at $z=2.43$ \citep[from the Horizon-AGN suite of simulations][]{Dubois2014,Peirani2017}. We then derive the \Lya forest absorption features from the emulated fields. 
Previous work followed a similar initial strategy to derive emission and absorption features of the IGM \citep{Villaescusa_Navarro_2013,Harrington-2021} . In this work, we improve on their work by a larger set of tests of the accuracy of the models and allowing for flexibility to allow the models to be used in an observational context.

We also explored the method to re-calibrate the equation-of-state to improve the emulated \Lya absorption for different gas physics, allowing the existing \LyNet to be used in vastly different conditions, only paying with a small additional training set. This allows us to directly affect the quantity instead of simply adjusting the mean transmitted flux. This is a powerful tool for a fast simulation of \Lya forest, with the potential of generalising a machine learning model to emulate different gas physics beyond the training set.
This work aims to provide a framework and a proof of concept for generating more accurate emulators of the gas hydrodynamics and the \Lya forest in particular.

%------------------------------------------------------------------------------------------
This paper is organised as follows. We discuss the physical process and the model we use for the \Lya absorption system in Section~\ref{sec:physical_model}. In Section~\ref{sec:method}, we motivate and describe the architecture of the neural network used for training, including the data transformations and the hyperparameters for training. We present the prediction results and accuracy benchmarks using \HnoAGN and \HDM in Section~\ref{sec:Results}. In Section~\ref{sec:transfer_learning}, we discuss the results using IllustrisTNG \citep{TNG100} as a test set and the transfer learning method to fine-tune the equation of state. We discuss the results and conclude in Section~\ref{sec:discussion}.

%------------------------------------------------------------------------------------------
\section{Physical model of the \Lya absorption}
\label{sec:physical_model}
In this section, we give a brief reminder of the fundamentals of the physical mechanisms that produce the observed \Lya forest. The details of the atomic physics underlying this derivation are given in earlier work by \cite{Meiksin_2009}. We focus here on the relevant physical quantities we seek to model with our framework.

We can break down the mechanisms behind the observed \Lya absorption features into a few components:  the intrinsic linewidth of the resonance of the \Lya transition, the thermal broadening owing to the small-scale random motions of the atoms, and the Doppler shift due to the bulk motions of the hydrogen clouds. 

The absorption strength of the broadband \Lya emission from one QSO allows us to infer the density of the neutral hydrogen at different distances along a line-of-sight. This absorption rate depends on the number of interactions. We can refer to it as opacity which is a function of the observed frequency $\nu_\mathrm{obs}$ as
\begin{equation}
\tau(D_\mathrm{QSO},\hat{u}, \nu_\mathrm{obs}) = \int_0^{s_{\text{QSO}}(D_\mathrm{QSO})}  \nHI(s,\hat{u}) \sigHI(\nu_\mathrm{obs},s,\hat{u}) \; \mathrm{d}s\, , 
\label{eq:opacity}
\end{equation}
where $\nHI$ is the number density of neutral atomic hydrogen, $\sigHI$ is the interaction cross-section of photons with the HI atoms. We note that $s$ is a physical proper distance. As such, $s_{\text{QSO}}(D_{\text{QSO}})$ is the physical proper distance corresponding to the comoving distance $D_{\text{QSO}}$ of the QSO on the considered line-of-sight.

To derive the expression for all these quantities, we first consider the situation in the rest frame of some atomic hydrogen in space. The atomic cross-section can be described by the Lorentz profile with the following expression
\begin{equation}
     \sigHI(\nu) = \frac{\pi e^2}{ m_e c}  f_{lu} L(\nu) =\frac{\pi e^2}{ m_e c}  f_{lu} \frac{\Gamma_{ul}/(4 \pi^2)}{(\nu - \nu_{lu})^2+(\Gamma_{ul}/( 4\pi)^2)}\, ,
\label{eq:cross_section_Loz}
\end{equation}
where $L(\nu)$ is the Lorentz profile,  $\nu_{lu}$ is the frequency of the transition from lower ($l$) to upper ($u$) energy level, $\Gamma_{ul}$ is the upper energy level damping width, $f_{lu}$ is the oscillator strength, $m_{e}$ and $e$ are electron mass and its charge, respectively. We note that the transition levels for the \Lya system are  $l=1$ and $u=2$, which gives $f_{12} = 0.4162$.

In addition to the probabilities of the fundamental transition, we have to consider two other components. The HI atoms have finite positive temperatures that induce their thermal motions, which for non-relativistic gas is well described by a Maxwell distribution. This effect is called Doppler broadening, and we denote the distribution by $G(\nu, T$). This type of broadening depends on the temperature of the gas, which is itself position-dependent. As HI clouds are in motion with respect to the photons emitted by the QSO, we have to apply a further Doppler boost to put the incoming photon in the rest frame of the cloud.

By taking the redshift of the photon emitter into account, this gives the relation between the frequency in the two frames to be
\begin{equation}
    \nu = \nu_\text{{obs}} (1+z)\, (1+\frac{v_{\text{z}}}{c})
\end{equation}
where $\nu$ is the central frequency, $\nu_\text{{obs}}$ the observed frequency,and $v_{\text{z}}$ is the line-of-sight velocity. Therefore the cross-section in this scenario is the convolution of the Lorentz and Doppler profile, 
\begin{equation}
  \sigHI(\nu, T) = \frac{\pi e^2}{m_e c} f_{12} (L * G)(\nu, T)  = \frac{\pi e^2}{m_e c} f_{12} V(\nu, T).
\end{equation}
The convolution of these two profiles is called a Voigt profile, which we write as $V(\nu, T)$:
\begin{multline}
    V(\nu, T) = \int_{\nu'=-\infty}^{\infty} \mathrm{d}\nu' \frac{\Gamma_{21}/(4 \pi^2)}{(\nu' - \nu_{12})^2+(\Gamma_{21}/( 4\pi)^2)} \\
    \frac{1}{\sqrt{\pi} \Delta \nu_{D}(T)} \exp\left[ - \left( \frac{\nu - \nu'}{\Delta \nu_D(T)}\right)^2\right]\, ,
     \label{eq:voigt_nonsimplified}
\end{multline} 
where
\begin{equation}
    \Delta \nu_D  = \frac{\nu_{12}}{c} \sqrt{\frac{2k_B T}{m_H}}
\end{equation}
is the Doppler width, which is the frequency broadening of the Lyman-$\alpha$ frequency due to the Doppler effect, and $\nu_{12}$ is the Lyman-$\alpha$ frequency.

We may rearrange the terms to simplify the expression by introducing $x = (\nu - \nu')/\Delta\nu_D(T)$, which is the frequency offset in the unit of $\Delta\nu_D$. We set $a = \Gamma_{21}/(4 \pi \Delta\nu_D)$ to be a ratio of the \Lya line width to the Doppler frequency width $\Delta \nu_D$.
We can now rewrite the equation \eqref{eq:voigt_nonsimplified} to be:
\begin{multline}
 V(\nu,T) = \\
 \int_{x'}\text{d}x' \left(  \frac{1}{\pi \Delta \nu_D} \right) \left( \frac{a}{(x(\nu)-x')^2+a^2} \right) \frac{1}{ \sqrt{\pi}\Delta \nu_{D}} \exp \left( - x'^2\right)\, .
\end{multline}
The Voigt profile then reads
\begin{equation}
 V(a,x) = \frac{a}{\pi^{3/2}\Delta \nu_D } \int^{\infty}_{-\infty} \frac{e^{-y^2 }  }{(x-y)^2 + a^2}dy\, .
\end{equation}
The normalisation of the Voigt profile is unity. We further transform the expression to express the Voigt profile $V(a,x)$ in terms of the  Voigt function $H$ as 
\begin{equation}
  V(a,x) = \frac{1}{\sqrt{\pi} \Delta \nu_D} H(a,x)\;.
\end{equation}
In the next section, we come back to the reason for using this function to increase computational efficiency.
This convention allows us to use the amplitude of the absorption, which yields
\begin{equation}
  H(a,x)=\frac{a}{\pi} \int_{-\infty}^{\infty} \frac{e^{-y^{2}} d y}{a^{2}+(x-y)^{2}}\,.
\end{equation}
Therefore the cross-section of the Lyman-$\alpha$ absorption in the frame of the observer is,
\begin{equation}
\sigHI(\nu_\mathrm{obs}) = \frac{\pi e^2}{ m_e c}  f_{12} \frac{H(a,x)}{\sqrt{\pi} \Delta \nu}\,.
\label{eq:cross section}
\end{equation}
With the cross-section in Equation~\eqref{eq:cross section}, we then compute the normalised absorption rate for Lyman-$\alpha$, also named the Lyman-$\alpha$ forest, as
\begin{equation}
  F(\lambda) = \exp\left[-\tau\left(\nu_\mathrm{obs}\right)\right].
  \label{eq:norm_flux}
\end{equation}
Thus, we have an explicit algorithm for computing the absorption features encoded by $F(\lambda)$. We detail the numerical implementation in Section~\ref{sec:mock_generation}. We note that we have not discussed the influence of the UV background, which directly affects the amount of neutral hydrogen \citep{Ionizing_UV}, the relation between the spin temperature and the gas temperature \citep{Liszt2001}, and possible relativistic effects \citep{Iri_2016}. Also, by considering the function $F(\lambda)$, we assume that the \Lya profile of the QSO is fully known since this is not the subject of this work. We only focus on how to emulate the quickest and the broadest possible effect of hydrodynamical simulation to obtain the function $F(\lambda)$.

\subsection{Benchmark Model : modified-FGPA}
\label{subsec:FGPA}

To assess the performance of our model to generate \Lya absorption, we employ a similar approach to the fluctuating Gunn-Peterson approximation \citep[FGPA,][]{Gunn1965,Weinberg1997} which is an analytical framework to estimate a mock \Lya from dark matter overdensity based on the idea that dark matter gravitational potential attracts baryonic matter \citep{Bi_1997}.
The approximation expresses as 
\begin{equation}
    \tau(z,\mvec{r}) \propto (1+\delta(\mvec{r}))^\beta,
\end{equation}
where $\tau$ is the IGM optical depth at redshift $z$ at a comoving position $\mvec{r}$, $\delta$ is the dark matter overdensity, and $\beta$ is derived from the slope of temperature-density power law relation, which relies on the equilibrium between HI collisional recombination and photoionisation.

We use a similar approximation for each individual fields, which yields the following relation for the neutral hydrogen density:
\begin{equation}
\label{eq:powerlaw_rho}
    \nHI^\text{mock} = \bar{n}_\text{HI} \left( \frac{\rho_\mathrm{DM}}{\bar{\rho}_\mathrm{DM}} \right)^{\alpha_{n}}\;,
\end{equation}
where $\bar{\rho}_\mathrm{DM}$ is dark matter average density, $n_\text{HI}^\text{mock}$ is an estimated neutral hydrogen number density, and $\bar{n}_\text{HI}$ is the coefficient of the power law.

And from the temperature-density power law relation, we can estimate a mock hydrogen temperature by using
\begin{equation}
\label{eq:powerlaw_T}
    T_\text{HI}^\text{mock} = \bar{T}_\text{HI} \left( \frac{\rho_\mathrm{DM}}{\bar{\rho}_\mathrm{DM}} \right)^{\alpha_{T}}
\end{equation}
where $ T^\text{mock}$ is an estimated hydrogen temperature and $\bar{T}_\text{HI}$ is the coefficient of the power law. The parameters of the dark matter from \HDM fitting with \HnoAGN hydrogen density and temperature shows in Table~\ref{tab:FGPA}.
From here, we refer to the benchmark model as modified-FGPA (mod-FGPA). Furthermore, to compare modified-FGPA to \LyNet, we apply the estimated density and temperature into the Equation~\eqref{eq:opacity}, which also includes the Voigt profile, and we assume that the line-of-sight velocity of dark matter and gas are approximately the same. 

We also note that this model is not strictly speaking the FGPA, but it employs the same power-law approximation for the relation between the gas-state fields and the dark matter. We expect the result of this test to be more optimistic than classical FGPA but worse than the one obtained with \LyNet because it is less flexible and it ignores the stochastic components of the gas.

\begin{table}[]
\caption{Fitted parameters of the power-law relation of dark matter and IGM parameters in Equation~\eqref{eq:powerlaw_rho} and \eqref{eq:powerlaw_T}. We use \HDM  dark matter overdensity and \HnoAGN gas density and temperature.}
\label{tab:FGPA}
\centering
\begin{tabular}{@{}lll@{}}
\toprule
\textbf{Gas Parameter ($X$)} & {$\mathbf{\bar{X}}$} & {$\mathbf{\alpha_X}$} \\ \midrule
Temperature   & 3.98      & 0.58            \\
Density       & -10.43    & 1.27            \\ \bottomrule
\end{tabular}
\end{table}

%--------------------------------------------------------------------
\section{Method}
\label{sec:method}
The model we use to calculate the absorption at different frequencies of the \Lya broad emission line needs three ingredients that characterise the physical state of the neutral hydrogen: the temperature, the gas density, and the line-of-sight gas velocity. Therefore, the main objective of this work is to construct the neural network which can produce these parameters for the computation of the model of the \Lya forest. We train \LyNet for each gas parameter separately to accommodate the optimisation and the prediction interpretation. This section discuses the details of the cosmological simulation used as the training set, the numerical approach for generating the mock \Lya absorption, and the data pre-processing procedure.

\begin{figure}
    \centering
    \includegraphics[width=\hsize ]{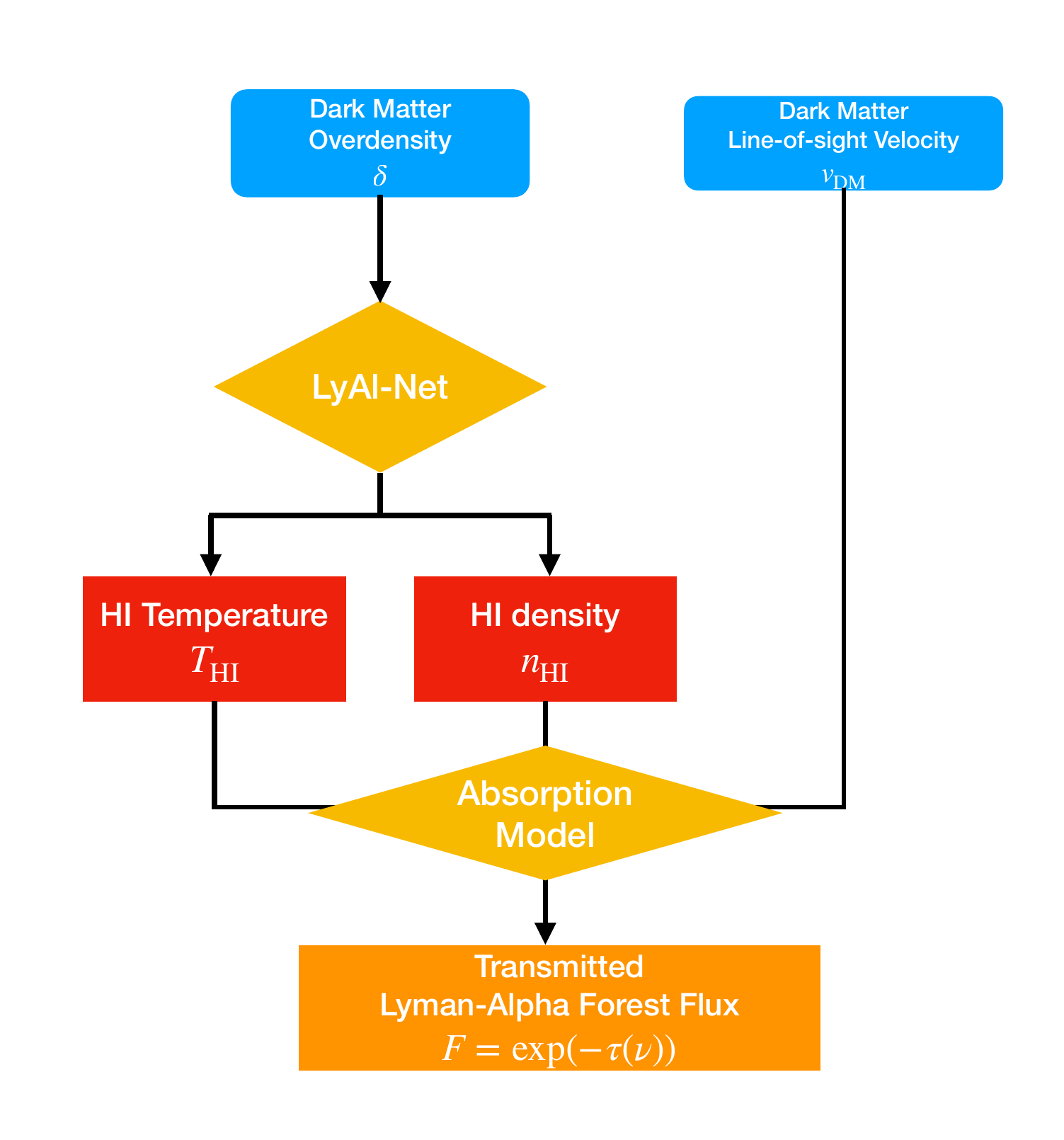}
    \caption{The flowchart illustrates a schematic of the \LyNet pipeline for the Lyman Alpha forest simulation. The flux absorption model (Equation~\eqref{eq:norm_flux}) requires three gas physical quantities, namely temperature, density, and the line-of-sight velocity. \ LyNet emulates only the gas temperature and the gas density of neutral atomic hydrogen. We use dark matter line-of-sight velocity of the same $N$-body simulation as an approximated gas line-of-sight velocity.}
    \label{fig:LyAl-Net_overview}
\end{figure}

\subsection{The simulation datasets}
\label{sec:simulation}

The simulation that we used is the Horizon-AGN simulation. It is a cosmological hydrodynamic simulation based on the adaptive mesh refinement code \ramses\citep{Teyssier2002_RAMSES}. The simulation contains $1024^3$ dark matter particles with a resolution of $M_\mathrm{DM,res}=8 \times 10^7\;\Msun$ and the volume span a cube of $100\Mpch$, which took 4 million CPU hours to complete with a redshift down to $z=1.2$. Horizon-AGN has been computed assuming a standard \LCDM cosmology and adopting cosmological parameters compatible with the data of the WMAP mission  \citep{2011WMAP7}: $\Omega_m = 0.272$, $\Omega_\Lambda = 0.728$, $\sigma_8=0.81$, $\Omega_b=0.045$, $H_0 = 70.4\;\mathrm{km s^{-1} Mpc^{-1}}$, and $n_s=0.967$. 
Furthermore, the simulation models the gas as ideal monatomic gas with adiabatic index $\gamma = 5/3$. The gas cools down via a metal-dependent model of hydrogen and helium photon emission \citep{Sutherland1993}. This allows the temperature of the gas to cool down to $10^4\;\mathrm{K}$. At the same time, the gas is heated via a uniform UV background following a model by \citet{Haardt1996} after redshift $z_\text{reion} = 10$. 
In this work, we used the \HnoAGN as a training set. It is the sibling simulation of Horizon-AGN with one major difference: the noAGN variant does not have AGN feedback due to the lack of black hole growth. The reason for using noAGN is that the computational time for the simulation was reduced by turning off the AGN feedback and emulating their effect by fine-tuning the UV background radiation to achieve the same gas properties \citep{Peirani_2022}. Another upside of using \HnoAGN is that we can study the effects and sensitivities of the \Lya forest model. We also used TNG100 simulation as a validation set to properly cross-check the results in Section~\ref{subsubsection:TNG-100:Transfer Learning}.

\subsection{Dataset generation for training and validation}
\label{sec:dataset_generation}

In this section, we describe the steps involved in creating the different datasets required for the training and validating the Machine Learning model. 

The phase space distribution of dark matter is sampled with particles in the cosmological simulations that we considered. For our emulation strategy, we require the densities to be on a regular grid mesh to facilitate the computations. We have assigned the particles to a $1024^3$ grid mesh, leading to a resolution of $\sim 98\kpch$ for each voxel. Several mass assignment schemes exist, among which the most used is the Cloud-In-Cell assignment \citep{Hockney1988}. Unfortunately, this may lead to mesh cells completely devoid of mass and prone to very poor sampling in cosmic voids, which is exactly where the signal of the \Lya forest is the most relevant \citep{Porqueres_2020}. We rely on the particle assignment developed by \cite{Colombi_2007} to avoid these problems. This assignment relies on an adaptive filter derived from Smooth Particle Hydrodynamics (SPH) kernel \citep{Monaghan1992}. 

We briefly remind the reader of the procedure here. The complete classical description of a particle system is given by the phase-space distribution,  $f(\mvec{r},\mvec{v}_L)$. For our computations, we typically require the moment of that distribution, which is for the velocity:
\begin{equation}
    \mvec{v}(\mvec{r}) = \frac{1}{\rho}\int \text{d}^3 \mvec{v}_L\; \mvec{v}_L f(\mvec{r},\mvec{v}_L),
\label{eq:Filter_velocity}
\end{equation}
where $f(\mvec{r},\mvec{v})$ is the phase space density of the dark matter particles.
The mass density $\rho(\mvec{r})$ is obtained similarly from the integral
\begin{equation}
    \rho(\mvec{r}) = \int \text{d}^3 \mvec{v}_L f(\mvec{r},\mvec{v}_L).
\label{eq:Filter_density}
\end{equation}
In practice, we compute  Equations~\eqref{eq:Filter_velocity} and \eqref{eq:Filter_density} into a mesh made of small cubic patch $\Delta r^3$, and approximate the resulting fields piecewise. We do these computations using an approach similar to smooth particle hydrodynamics \citep{Monaghan1992} to reduce shot noise. We can represent each particle as a smooth cloud with a finite size based on their local density, which depends on the distances between neighbouring particles. Generally, we define the cloud size as $2R_\mathrm{SPH}$ and the number of neighbouring particles as $N_\mathrm{SPH}$. The algorithm we use preserves the conservation of mass and momentum by an adequate choice of mesh node weighing, as detailed below.

To interpolate particles into the grid site, we use the following equation
\begin{equation}
\label{eq:A SPH}
\tilde{A}(i, j, k)=\frac{1}{\left[R_{\mathrm{SPH}}(i, j, k)\right]^{3}}\left[\sum_{l=1}^{N_{\mathrm{X}}-1} A_{l} W_{l} \mathcal{S}\left(\frac{d_{l}}{R_{\mathrm{SPH}}(i, j, k)}\right)\right] .
\end{equation}
This equation provides a weighted sum of a quantity $A_l$ carried by a particle $l$. Each neighbouring particle is multiplied by $\mathcal{S}(x)$, an SPH kernel where $x$ is the relative distance by calculating the ratio of the distance $d_l$ of a particle $l$ from the grid mesh node $(i,j,k)$ and $R_{\mathrm{SPH}}(i,j,k)$. They are weighted using the $W_l$ factor such that the total sum for each grid site is equal to unity. This constraint yields the following identity for the weight of each particle:
\begin{equation}
W_{l}=1 / S_{l}=\left[\sum_{i, j, k} \frac{1}{\left[R_{\mathrm{SPH}}(i, j, k)\right]^{3}} \mathcal{S}\left(\frac{d_{l}}{R_{\mathrm{SPH}}(i, j, k)}\right)\right]^{-1}\;.
\end{equation}
For the particular case of assigning the mass to the grid, we end up with a mass per mesh node of the grid. We divide by the volume corresponding to this mesh node to obtain the mass density. This yields:
\begin{equation}
    \Tilde{\rho}(i,j,k) = \Tilde{m}(i,j,k)/\Delta r^3\;.
\end{equation}
Similarly, the interpolated velocity is obtained following the equation~\eqref{eq:Filter_velocity} which gives the mesh the following identity:
\begin{equation}
    \Tilde{v}(i,j,k)=\Tilde{p}(i,j,k)/\Tilde{m}(i,j,k)\;,
\end{equation}
where $\Tilde{p}$ is the interpolated momentum of the cell.

Unlike dark matter, the gas simulation from \ramses has an adaptive mesh refinement structure. Each raw gas cells have a different comoving size. The largest cells have an extent of $0.0977\Mpch$, while the smallest cells have 1/32th of this length. We homogenised the resolution by assigning the gas cell based on its coordinate to the nearest grid, similar to the nearest grid point scheme. Some of the mesh cells may contain multiple sub-cell, e.g. two $L_\text{largest}/4$ cells and one $L_\text{largest}/2$ cells. We used the same weight regardless of the length to average the quantities, which is a sufficient approximation given the scale we are working on with the \Lya system. The algorithm can easily be expressed as  
\begin{equation}
    {X_{cell}}(i,j,k) = \frac{1}{N}\sum_{i=1}^N x_i,
\end{equation}
where $X_{cell}$ is the target parameter at coordinate $(i,j,k)$ and $x_i$ is the sub-cell value of the target parameter.

\subsection{Mock Lyman-$\alpha$ absorption generation}
\label{sec:mock_generation}

We have discussed the physics of the \Lya forest earlier in Section~\ref{sec:physical_model}. Let us now move to the numerical approach to calculate the normalised flux from the neutral hydrogen parameters.
The opacity of the gas as a function of the observed frequency of a given line-of-sight defined in Equation~\eqref{eq:opacity} is in the continuous form. However, since the dataset is in a grid mesh structure, we have replaced Equation~\eqref{eq:opacity} with a Riemann sum to calculate the integral. The opacity now reads as
\begin{equation}
\tau_{\vec{a},k} = \sum_{j=0}^{j=N} n_{\mathrm{HI},\vec{a},l_j} \sigHI(\nu_{\mathrm{obs},j},\vec{a},j)\times \delta l\;, 
\label{eq: opacity discre}
\end{equation}
where $j$ is the index of the position with respect to the line-of-sight, $l_j = j \times \delta l$ is the proper distance from an observer to a source, $\vec{a}$ is the unit vector which is perpendicular to a source plane, $\delta l$ is the physical length of each cell width, which in our case is $\delta l= 0.098 \Mpch$.
The Voigt function is expensive to compute numerically. We opted to rely on the Faddeeva function, which is precomputed as part of SciPy \citep{SciPy}. We refer the reader to Appendix~\ref{app:voigt_profile} for the relation between this function and the actual computation we do. We then compute the normalised absorption flux $F(\nu)$ by following Equation~\eqref{eq:norm_flux}. We do not consider the exact shape of the broadband emission Lyman-$\alpha$ line of the mock QSO itself, which has to be estimated as usual in the observational context.

Converting the spatial resolution of the simulation to spectrometer resolution gives a resolution of $R = \frac{\langle \Delta \lambda \rangle}{\langle\lambda\rangle} \approx 30000 $. This resolution is much higher than what is possible to achieve with current cosmological surveys. For example, WEAVE-QSO has $R=5000$ and $R=20000$ in low and high-resolution mode \citep{Dalton2014}, and SDSS IV  (BOSS and eBOSS) has $R=2000$. This allows us to have an error budget in the simulation compared to the real surveys.
Of course, a high resolution does not mean the physics of the simulation is accurate, nor does it contains all of the baryonic physics at that resolution. We will consider the reliability of our model by comparing results obtained with other simulations in Section~\ref{sec:transfer_learning}.
\begin{figure}
   \centering
   \includegraphics[width=\hsize ]{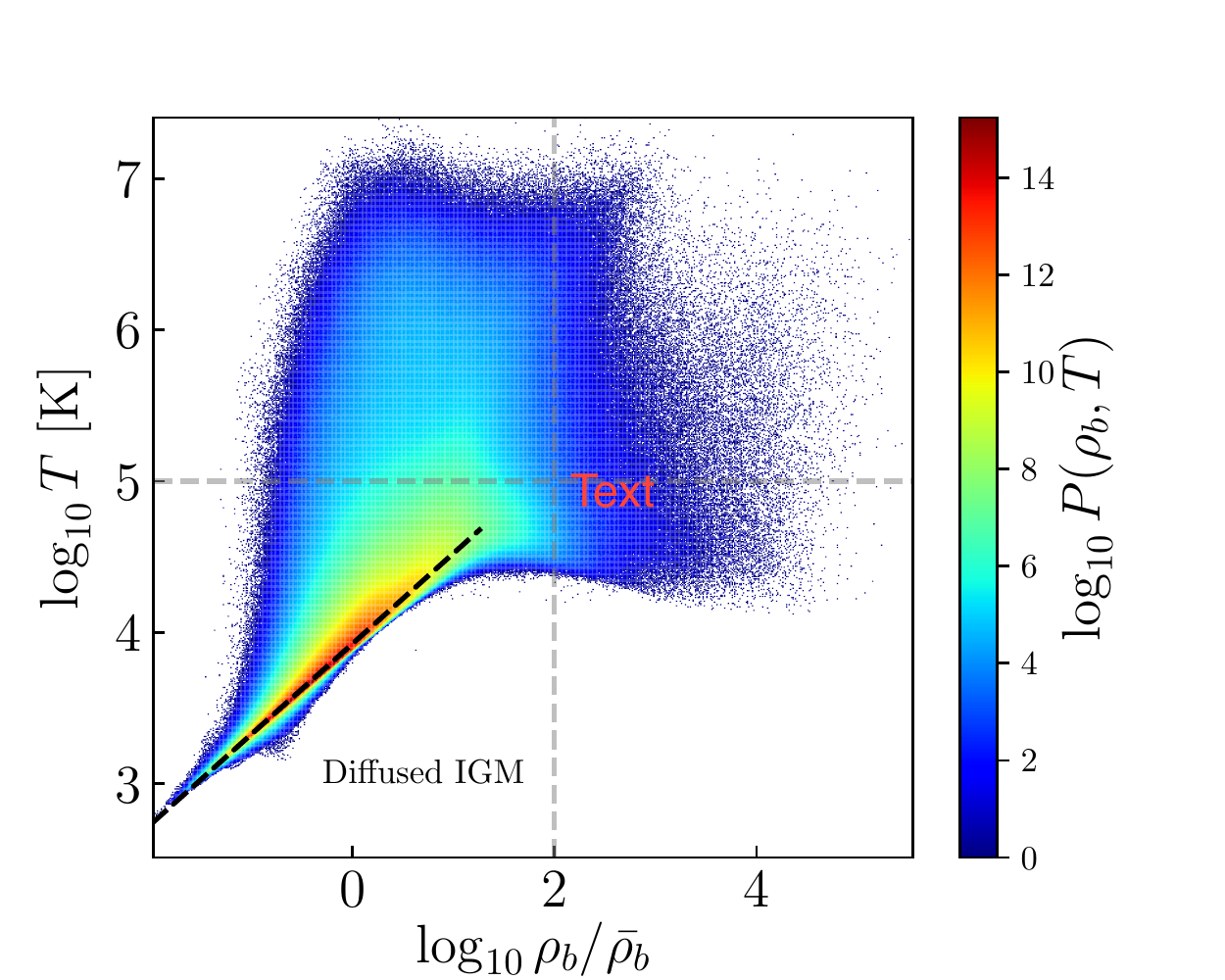} 
      \caption{A Density-Temperature ($n-T$) diagram of hydrogen density vs temperature in the \HnoAGN simulation at $z \approx 2.4$. The black dashed line illustrates the fitted strong power-law relation within the diffused IGM regime, which is the main contribution of \Lya forest, corresponding to ~99.5~\% of the total volume and ~76.7~\% of the total hydrogen mass. Light grey dashed line separates the different baryonic regimes. We provide the density plot in this phase diagram by colouring each point with a log scale on the right-hand side.}
         \label{fig:phase_diagram_Horiz}
\end{figure}
\subsection{Pre-transformation of the fields before training}
\begin{figure*}
   \includegraphics[width=\hsize]{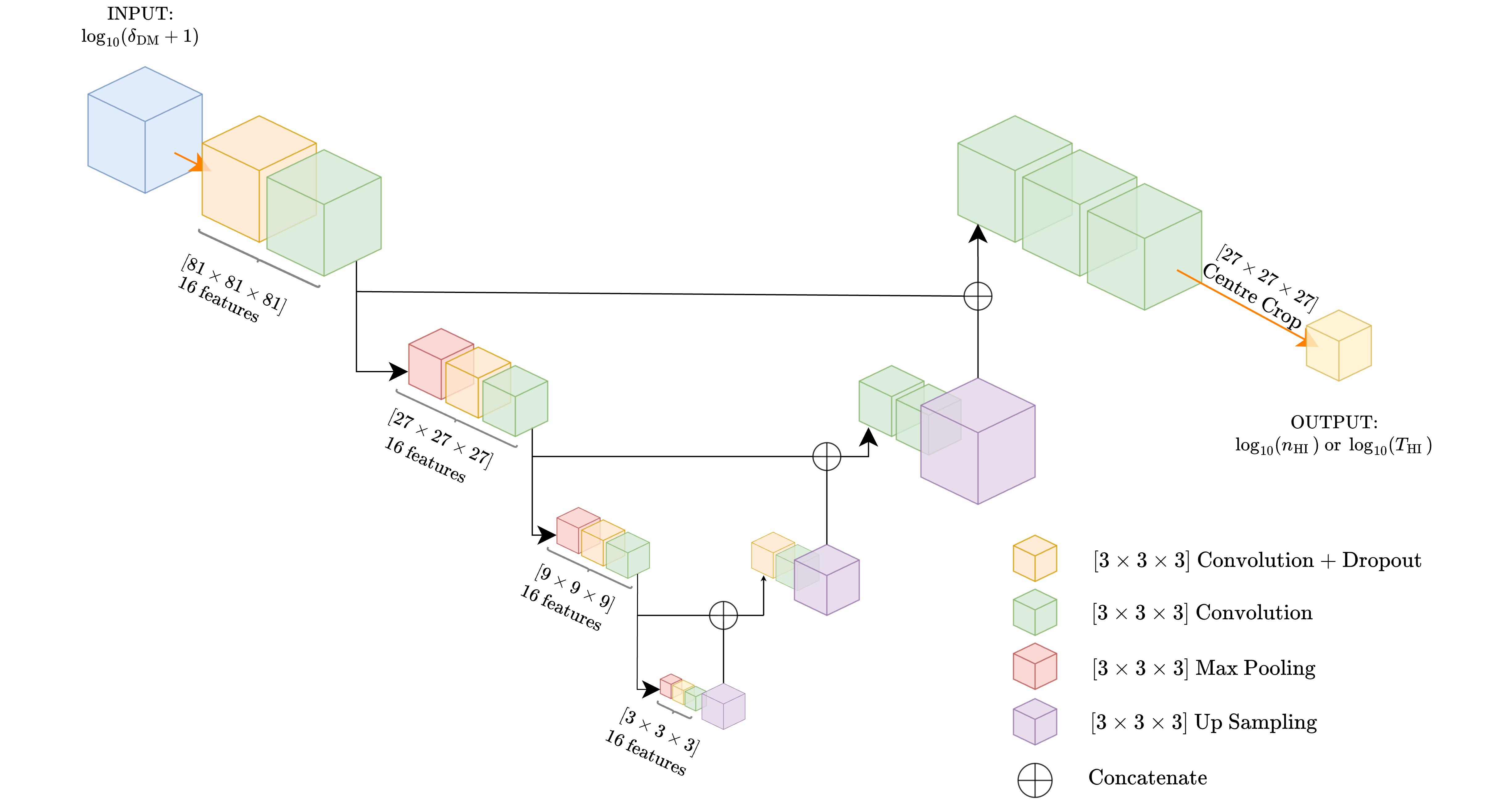}
    \caption{We show here the schematic of the \LyNet architecture. The input of the network is the transformed dark matter field, $\log_{10} \left( \delta_{DM}+1 \right)$, with the size of $81\times 81\times 81$ voxels. The output is the targeted gas parameter (temperature or gas density) with the size of $27\times27\times27$ pixels. We crop the volume at its centre from the last convolutional layer. The arrow represents the direction of the data where each colour represents a different operation with $3\times 3\times 3$ kernel size. We note that all of the convolution layers extract 16 feature maps. The model is trained to predict each gas parameter separately for the simplicity of optimisation and interpretation.}
    \label{fig:Unet}
\end{figure*}

The dark matter density field is usually described in terms of the density contrast defined as 
\begin{equation}
\delta_\mathrm{DM}= \frac{\rho_\mathrm{DM}}{\bar{\rho}_\mathrm{DM}} - 1\;,
\end{equation}
where $\bar{\rho}_\mathrm{DM}$ is dark matter average density. The 1D distribution  of $\delta_\mathrm{DM}$ is heavily skewed with a range from -1 to infinity by construction. This is not necessarily the most convenient for emulators to manipulate. Furthermore, given enough flexibility, the skewed dataset does not affect the neural network. However, this transformation benefits several things: it compresses the dynamic range, allowing a scale-invariant process, and should help the training speed and reduce the required sample size to achieve a given accuracy. 

Therefore, we will provide the transformed density $A_\mathrm{\rho}$ derived from the density contrast as 
\begin{equation}
    A_\mathrm{\rho}(\rho_\mathrm{DM}) = \log_{10}\left(\frac{\rho_\mathrm{DM}}{\bar{\rho}_\mathrm{DM}} \right)=\log_{10}(1+\delta_\mathrm{DM})\;.
\label{eq:log_dm}
\end{equation}
Similarly, we applied a decimal logarithmic transformation for the density and temperature of neutral hydrogen.
\begin{equation}
A_{n,\mathrm{HI}} = \log_{10} \frac{\nHI}{1 \text{g cm}^{-3}}\;,
\end{equation}
and
\begin{equation}
A_{T,\mathrm{HI}} = \log_{10} \frac{\TempHI}{1\text{ K}}\;.
\end{equation}

The choice of this transformation for density and temperature is not only for the coherence of the scale used between the input and prediction. This transformation should facilitate our interpretation of the differences in the dark matter distribution in the prediction analysis. Most importantly, following the suggestion that the relation between density and temperature is a power law in a diffuse IGM regime, which is typically located in the void regions \citep{Hui1997,Schaye_1999,Meiksin_2009}, 
this temperature-density relation expresses as

\begin{equation}
    T = T_0 \left(\frac{\rho_b}{\bar{\rho}_b}\right)^{\gamma-1},
    \label{eq:power-law_temp-rho}
\end{equation}
where $\gamma$ is the adiabatic coefficient, $T_0$ is the temperature at the mean density, $\rho_b$ is the density of baryonic matter, and $\bar{\rho}_b$ is the mean baryonic density.
Since we are interested in neutral hydrogen, for simplicity, we assume $\frac{\rho_b}{\bar{\rho}_b} \approx \frac{{n}_\mathrm{HI}}{\bar{n}_\mathrm{HI}}$.  Hence, a log transformation of this relation becomes a linear relation expressed as
\begin{equation}
    A_{T,\mathrm{HI}} =  A_{\bar{T},\mathrm{HI}} + (\gamma-1) \left(A_{n,\mathrm{HI}} - A_{\bar{n},\mathrm{HI}} \right) \;.
\end{equation} 
We show a density-temperature plot in Figure~\ref{fig:phase_diagram_Horiz}, and we highlight the diffuse IGM region using the threshold of the temperature is $T > 10^ 5$ K  and for the density is $n_\mathrm{H} < 10^{-4}(1+z)\;\text{cm}^{-3}$ \citep{Martizzi_2019}.
This gas phase contains the majority of baryonic density, corresponding to ~99.5~\% of the total volume and ~76.7~\% of the total hydrogen mass. Atomic hydrogen density is also embedded within this field, therefore, carrying the most important information for a \Lya forest.
For dark matter and gas velocities, we do not re-scale for the neural network training since they are similar in magnitude. The hope is that a neural network can construct a phenomenological model that can relate corrections to this mean relation.

\subsection{\LyNet Architecture}
\label{sec:architecture}
This section discusses the neural network architecture used for this project and its advantage. Then we will discuss the training process, including the size of the training sample, the choice of the loss function, and the details of the architecture of \LyNet. 
For this work, we chose a U-net architecture for the neural network, firstly introduced by \cite{Ronneberger2015} for biomedical image segmentation. 
A U-net is based on the convolutional neural network \citep[CNN, ][]{oshea2015}, which can extract and create feature mapping similar to how the human visual cortex works. It is well known for its pattern recognition ability, spatial invariance, and scale invariance. This architecture is much faster than the traditional artificial neural network \citep[ANNs, ][]{oshea2015}, such as multilayer perceptron. It has become an invaluable technique for solving various problems, especially image-focus problems. The significant difference between CNNs and ANNs consists of the convolutional layers. These layers allow the machine to perform convolutions with kernels and leverage on three crucial points: sparse interactions, parameter sharing, and equivariant representations \citep{GoodBengCour16}. The U-net architecture allows us to extract and summarise the information on a small scale while retaining large-scale information thanks to network skipping. With the knowledge that hydrogen gas is the tracer of dark matter and vice versa, the U-net becomes a practical approach to generating the gas fields in a short amount of time compared to the typical hydrodynamic simulation, which also has been used by \cite{wadekar2020hinet} and \cite{Bernardini_2021}. 

The schematic of the \LyNet is shown in Figure~\ref{fig:Unet}. It comprises three major steps: the convolution side (on the left), the bottleneck (at the bottom), and the up-sampling side (on the right). We chose the $3\times 3\times 3$ convolution kernel for each convolution step while maintaining the dimension using 'SAME' padding and adopting a dropout layer with a rate value of $0.2$.\footnote{A convolution kernel with 'SAME' padding adds zeros around the input, such as the output as the same dimensions as the input.}. The $3\times 3\times3$ max-pooling is also used to reduce the dimension for each contracting step.\footnote{Max-pooling operates by extracting the maximum value within the $3\times 3\times 3$ kernel, scanning in the tensor.} After the bottleneck phase, we used the $3\times 3\times3$ up-sampling layer followed by the same configuration of convolution. For each convolution layer of the up-sampling phase, we used 16 filters applied on the input concatenation and the bypass from the contracting side. 

The overall dimension of the input is $81\times81\times 81$ pixels, and the output is $27\times27\times27$ pixels, which is cropped to prevent the possible edge effect caused by the convolution. 
Thus, we need to tile the output into the larger cube to obtain the full simulation from the output, which will be explained later in Section~\ref{subsec:tiling}.
We note that this architecture is not optimal, and a better architecture can exist to provide an equivalent or better result with the smaller training parameters.

\subsubsection{Loss Function} 
We choose a mean squared error (MSE) as the loss function. The main advantage of using this loss function is that it is more stable compared to GAN loss \citep{Nguyen_2021}. It also allows a better understanding of what the neural network learns, therefore, better interpretability. It takes a form as

\begin{equation}
    \text{MSE }(Y_{i},\hat {Y_{i}})={\frac {1}{n}}\sum _{i=1}^{n}(Y_{i}-{\hat {Y_{i}}})^{2}\;,
\end{equation}
where $n$ is the batch size, $Y_i$ is the prediction, and $\hat{Y_{i}}$ is the true value.

We note that this loss is applied to the transformed variables ($A_X$), and it only calculates the loss of the central part with $27\times27\times27$ pixels. For the velocity field, this stays linear. For the other quantities, as  $T_{\text{HI}}$ and $n_{\text{HI}}$, owing to the log transformation, the effective loss function is the mean squared log error (MSLE) which reads
\begin{equation}
    \text{MSLE}(Y_{i},\hat {Y_{i}}) ={\frac {1}{n}}\sum _{i=1}^{n}\left(\log{ \frac{Y_{i}}{{\hat {Y_{i}}}}}\right)^{2}\;.
\end{equation} 
This loss weighs high and low values equally and eases the training process. The MSLE loss punishes the underestimation more than the overestimation. It is a two-sided choice, both carrying a risk and an advantage. Indeed it can lead to bias in the mean response, as we will investigate later. Doing so also avoids being overly sensitive to the high-value excursion, which is irrelevant for the \Lya forest. As long as the prediction is in a sensible window of error and the baryonic observation's resolution, this loss function should not be the primary concern. We will discuss prediction bias and error later in Section~\ref{sec:Results}.
%%%%%%%%%%%%%%%%%%%%%%%%%%%%%%%%%%%%%%%%%%%%%%%%%%%%%%%%%%%%%%%%%
\subsubsection{Training Process}
\label{sec:training}
We train \LyNet to predict each gas parameter separately because we can tweak the training hyper-parameters and interpret the prediction results independently from the other fields. 
We used \HnoAGN simulation as the training set, which contains both dark matter and gas components. It spans a volume of $(100\Mpch)^3$, with a resolution of  $1024$ pixels on each side. This snapshot has the redshift range from $z=2.44$ to $z=2.33$. Since we were limited to a single full simulation, we randomly selected 5120 sub-boxes, with each sub-box having a size of $81^3$ voxels or $\sim( 7.9\;\Mpch)^3$ in volume, which is equivalent to $\sim 9 \%$ of the total simulation volume. The data sampling also allows us to use repeated samples and overlapping of the training set, which encourages the network to learn translational invariance. The size of the sub-box is intended to fit the GPU memory and encourage the neural network to learn the information at a small scale. We note that a physically larger kernel size could improve the solution by better capturing the transfer of power from large to small scales. We will discuss this further in Section~\ref{sec:discussion}.

We also perform rotational augmentation \footnote{To increase the diversity of the dataset from the existing dataset} in all three axes, except the line-of-sight velocity which is rotated solely around the z-axis.
The rotation is required for the trained neural network to obtain the rotational invariance \citep{He_2019,Kaushal2021,Kodi_Ramanah_2019}, reflecting the physical process involved in the prediction. We expect this process to make the emulator robust. This property will be useful for transfer learning for a different dark matter simulation. 

Table~\ref{tab:hyper_param} shows summarised details of the parameters of the training loop that we used, which includes the number of epochs and the learning rate. We used a batch size of 64 with Adam Optimiser \citep{AdamOpti}. We note that the hyper-parameters of the \LyNet were obtained via numerical experimentation to ensure a repeatable and quick convergence.

\begin{table}
\caption{The table summarises the hyper-parameters used for \LyNet training with \HnoAGN.}
\begin{tabular}{@{}lcc@{}}
\toprule
\textbf{Target HI Parameter} & \textbf{Learning Rate} & \textbf{Number of Epochs} \\ \midrule
Density                       & 0.002                     & 1000                    \\
Temperature                   & 0.002                     & 1000                    \\
Velocity                      & 0.003                     & 100                    \\ \bottomrule
\end{tabular}

\label{tab:hyper_param}
\end{table}
\subsubsection{Tiling Algorithm}
\label{subsec:tiling}
As mentioned, the simulation size is discretised into $1024^3$ voxels. Since \LyNet produces emulated fields of only $27^3$ voxels from an $81^3$ voxels input, we have to tile emulated elementary cubes to achieve the desired volume before assessing the accuracy and quality of the prediction model. 
We implemented a tiling algorithm by applying a $27$ pixels sliding window, horizontally scanning from the XY-plane 38 times per row, column, and layer (Z direction). We describe with pseudo-code the algorithm in Algorithm~\ref{alg:tiling}. The total tiled volume consists of $38 \times 38 \times 38$  elementary cubes (equivalent to $1026^3$ voxels), which is then trimmed to obtain the $1024^3$ voxels. The construction of the complete volume took $\sim$ 8 minutes on Nvidia Tesla V100 and Intel(R) Xeon(R) Gold 6230, 40 cores. This cost can be further reduced by re-arranging some operations directly on the GPU to alleviate GPU-to-CPU data transfer. We postpone this optimisation for later, which could be useful for a fully differentiable model of the \Lya forest.

\section{Results on \HnoAGN }
\label{sec:Results}
\begin{figure*}
   \includegraphics[width=\hsize]{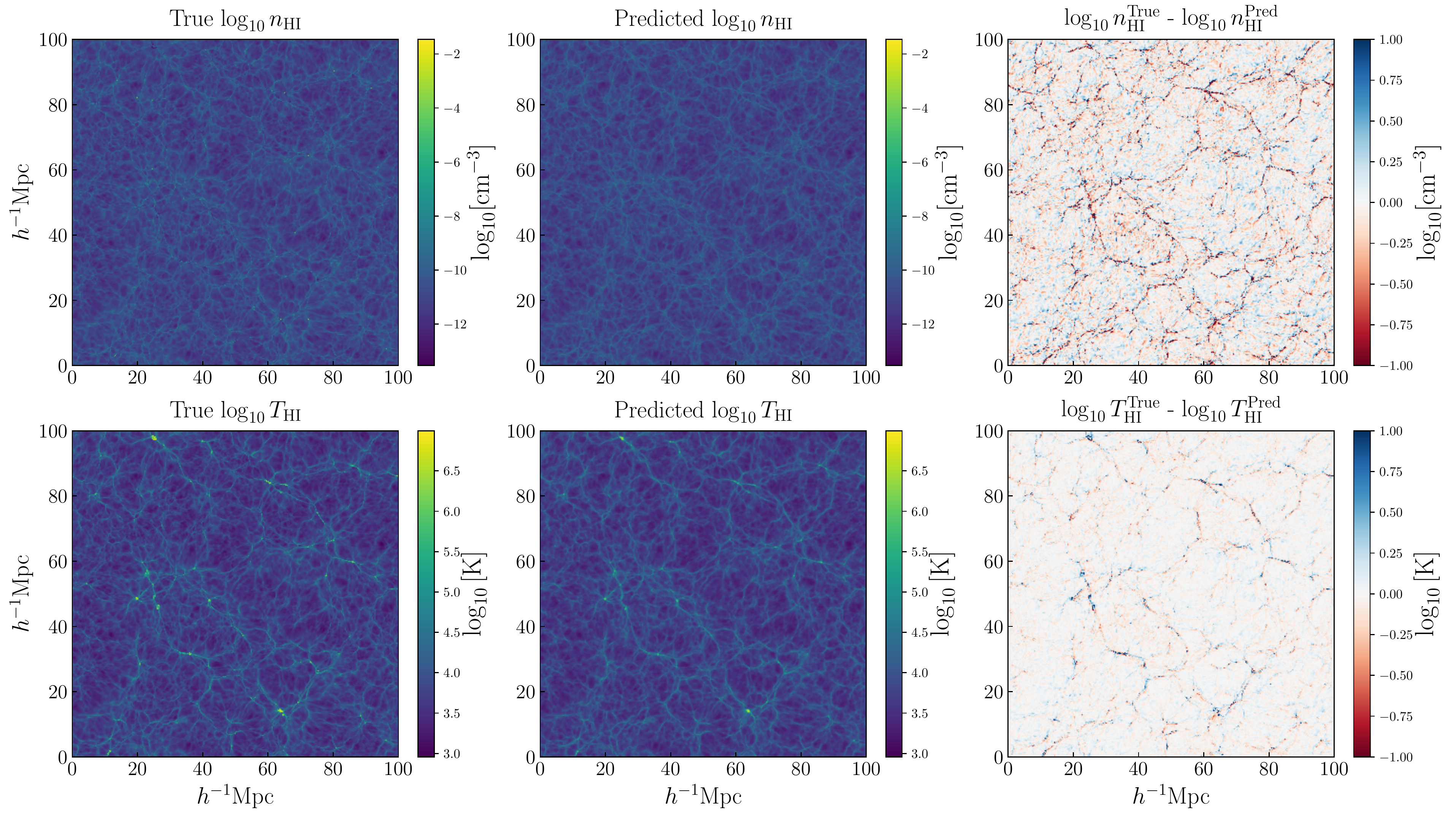}
      \caption{Comparisons of sample slices for different hydrodynamic quantities in the decimal logarithm scale.  Left column: fields directly extracted from \HnoAGN.  Middle column:  fields obtained from the application of \LyNet to the dark matter field. Right column: the difference between the leftmost panel and the middle panel. The atomic hydrogen number density (temperature, respectively) is presented on the top row (bottom row, respectively).
      } %
         \label{fig:Pred_sample}
\end{figure*}

In the following, we present the metrics we use to assess the quality of the emulated fields in Section~\ref{sec:quality}. Then we consider in turn each of the physical fields in the following section: the number density of atomic neutral hydrogen in Section~\ref{sec:result-nHI}, the temperature in Section~\ref{sec:result-temp} and the gas bulk velocity in Section~\ref{sec:result-vlos}. 
We emphasise that we train \LyNet with \HnoAGN, and use \HDM as a validation set to see how the model responds to different baryonic feedbacks. We also refer to the term "predicted fields" and "emulated fields" interchangeably.
\subsection{Prediction Quality Assessment}
\label{sec:quality}

We consider several metrics to assess the degree of trust that we have on \LyNet to predict physical fields. We will first consider the one-point statistics, i.e. the relation between the true and the predicted field at a given point in space. The second statistic that we will consider is the two-point correlation function which will provide insight into the reproduction of the mean spatial structure of the two respective fields.  

\subsubsection{One-point Statistics}
We consider several metrics to assess the degree of trust we may have on \LyNet to predict physical fields. We will first consider the one-point statistics, i.e. the relation between the true and the predicted field at a given point in space. The second statistic that we will consider is the two-point correlation function which will provide insight into the reproduction of the mean spatial structure of the two respective fields.  

\subsubsection{One-point Statistics}
We quantify the emulator's performance by considering the conditional probability distribution function of the predicted value $X_\text{Pred}$ of the field $X$ given the true value $X_\text{true}$ of that same $X$ from the simulation. This conditional probability is obtained from Bayes' theorem directly, which is expressed as
\begin{equation}
\label{eq:condition_prob}
    P(X_{\text{Pred}}|X_{\text{True}}) = \frac{ P(X_{\text{Pred}},X_{\text{True}})}{P(X_\text{True})},
\end{equation}
where $P(X_{\text{Pred}}|X_{\text{True}})$ is the probability of the predicted value of $X$ given the true value of $X=X_\text{Pred}$,  $P(X_{\text{Pred}},X_{\text{True}})$ is the joint probability of $X=X_{\text{Pred}}$ and $X=X_{\text{True}}$, and $P(X_{\text{True}})$ the probability of $X_{\text{True}}$.

We have estimated the joint probability distribution function for the predicted and true values $P(X_{\text{Pred}},X_{\text{True}})$ and $P(X_\text{True})$ by using the kernel density estimation \citep[KDE, ][]{Silverman1986,Scott2015}. It is a technique to infer a continuous probability density function (PDF) from a finite sample by smoothing the distribution using a weighted kernel function. Owing to the computational cost of this estimator, we built them on a reduced number of voxels. We, therefore, use approximately $50 \%$ of the total volume to calculate the KDE estimator. The full details are indicated in Appendix~\ref{app:KDE}.

%%%%%%%%%%%%%%%%%%%%%%%%%%%%%%%%%%%%%%%%%%%%%%%%%%%%%%%%%%%%%%%
\subsubsection{Two-point Statistics}
The two-point correlation function $\xi (r)$ is defined as 
\begin{equation}
\xi ({ \lvert\textbf{r} \rvert}) = \langle \delta_A(\textbf{r}') \delta_B(\textbf{r}'+\textbf{r} )\rangle\;.
\end{equation}
Compared to a homogeneous distribution, this measures the excess probability of finding two-point objects, field fluctuations, separated by a distance $\lvert \textbf{r} \rvert$. These objects or fields may be ,for example, galaxies or gas quantities. The power spectrum is the image of the correlation function in the spatial frequency domain. It is defined as
\begin{equation}
P(\lvert \textbf{k} \rvert) =  \int \mathrm{d}^3 \textbf{r}\; \xi(r) e^{-i \textbf{k} \cdot \textbf{r}}.
\end{equation}
The metrics that we used to assess both magnitudes and spatial information of the emulated fields are the transfer function $T_X(k)$ and the cross-correlation function $r_X(k)$. 

The transfer function $T_X(k)$, $k$ being the spatial comoving wave numbers, is a  convenient way to represent the departure of the predicted field to the original correlation structure of the true field $X$ for each $k$ scale. This function is commonly used in literature as a benchmark of the performance of an emulator or an approximation \citep[e.g.][]{Bardeen_1986,Leclercq_2013,Vlah2016,He_2019,Dai2021}. The transfer function is defined as
\begin{equation}
    T_X(k) = \sqrt{\frac{P_\text{pred,X}(k)}{P_\text{true,X}(k)}}\;,
\end{equation}
where $P_\mathrm{pred}(k)$ is the power spectrum of the emulated field, and $P_\mathrm{true}(k)$ is the reference or true field. 

Finally, we consider the correlation rate $r_X(k)$ to check the linear correspondence 
\begin{equation}
    r_X(k)  =  \frac{P_{\text{true,X} \times \text{pred,X}}(k)}{\sqrt{P_\text{true,X}(k)P_\text{pred,X}(k)}}\;.
\end{equation}

To compute the power spectrum of the different scalar fields considered in this work, we rely on the use of \texttt{nbodykit} \citep{nbodykit}. For the prediction quality assessment of \HnoAGN, we decide to include the trained portions of the dataset in the power spectra calculation because only $\leq 9 \%$\footnote{The data sampling algorithm for a training set allows repeated sampling.} of the simulation volume was used to train \LyNet. The main reason is to avoid an irregular volume for power spectra computation, and the volume should be small enough compared to the rest of the validation volume. 

\subsection{Neutral hydrogen Density (\nHI)}
\label{sec:result-nHI}

\begin{figure}
   \includegraphics[width=\hsize]{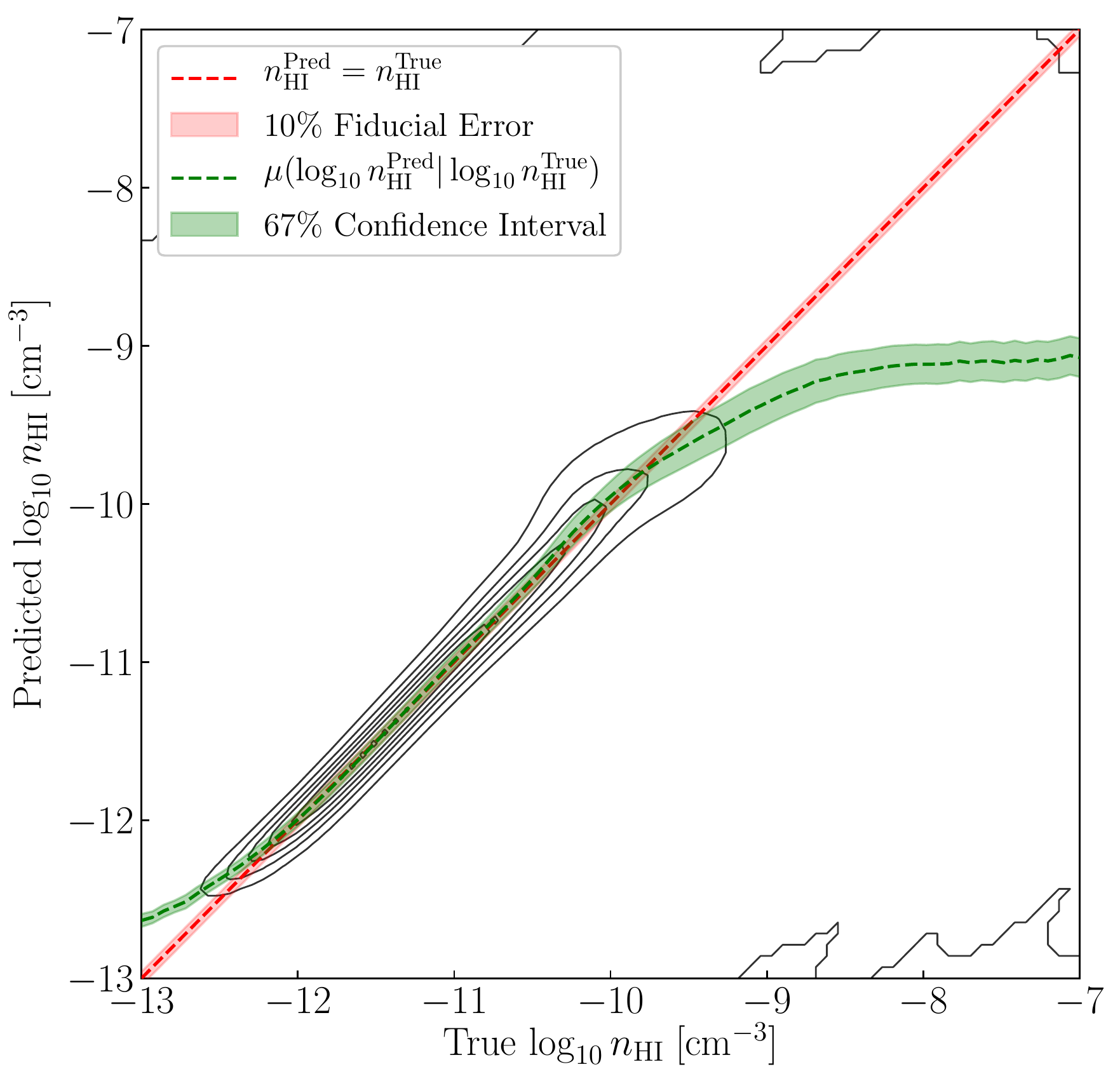}
     \caption{The conditional probability of the predicted decimal logarithm of density \nHI given the true density of the \HnoAGN and their corresponding marginal distribution where the black solid contour lines represent different levels of probabilities (see Equation~\eqref{eq:condition_prob}). The diagonal red-dashed line indicates the unbiased relation between the predicted and true density alongside a 10\% fiducial error budget. The green dashed green line indicates the mean of the conditional probability distribution alongside the 67\% confidence interval. We note a good agreement over several orders of magnitude of the predicted gas density from $1 \times 10^{-12}\,\text{cm}^{-3}$ and $6 \times 10^{-10} \,\text{cm}^{-3}$. A significant saturation occurs for very low density at $\nHI \le 10^{-12}~\text{cm}^{-3}$.}
         \label{fig:posterior_nHI}
\end{figure}
Among the different quantities required to compute the absorption rate in the Lyman-$\alpha$ forest, the atomic hydrogen number density is most critical to obtain correctly since the neutral hydrogen clouds are the absorber in the \Lya forest process. In the top row, Figure \ref{fig:Pred_sample} shows a slice of the atomic hydrogen number density \nHI from the \HnoAGN prediction from dark matter overdensity and the (logarithmic) difference between the ground truth and prediction. Visually, the structures of both slices are in good agreement. We note no apparent defects caused by \LyNet nor edge effects induced by our tiling procedure to cover the entire volume except for spots where a high relative error is present, typically correlated with high-density regions and where \Lya absorptions are saturated. 

\begin{figure}
   \includegraphics[width=\hsize]{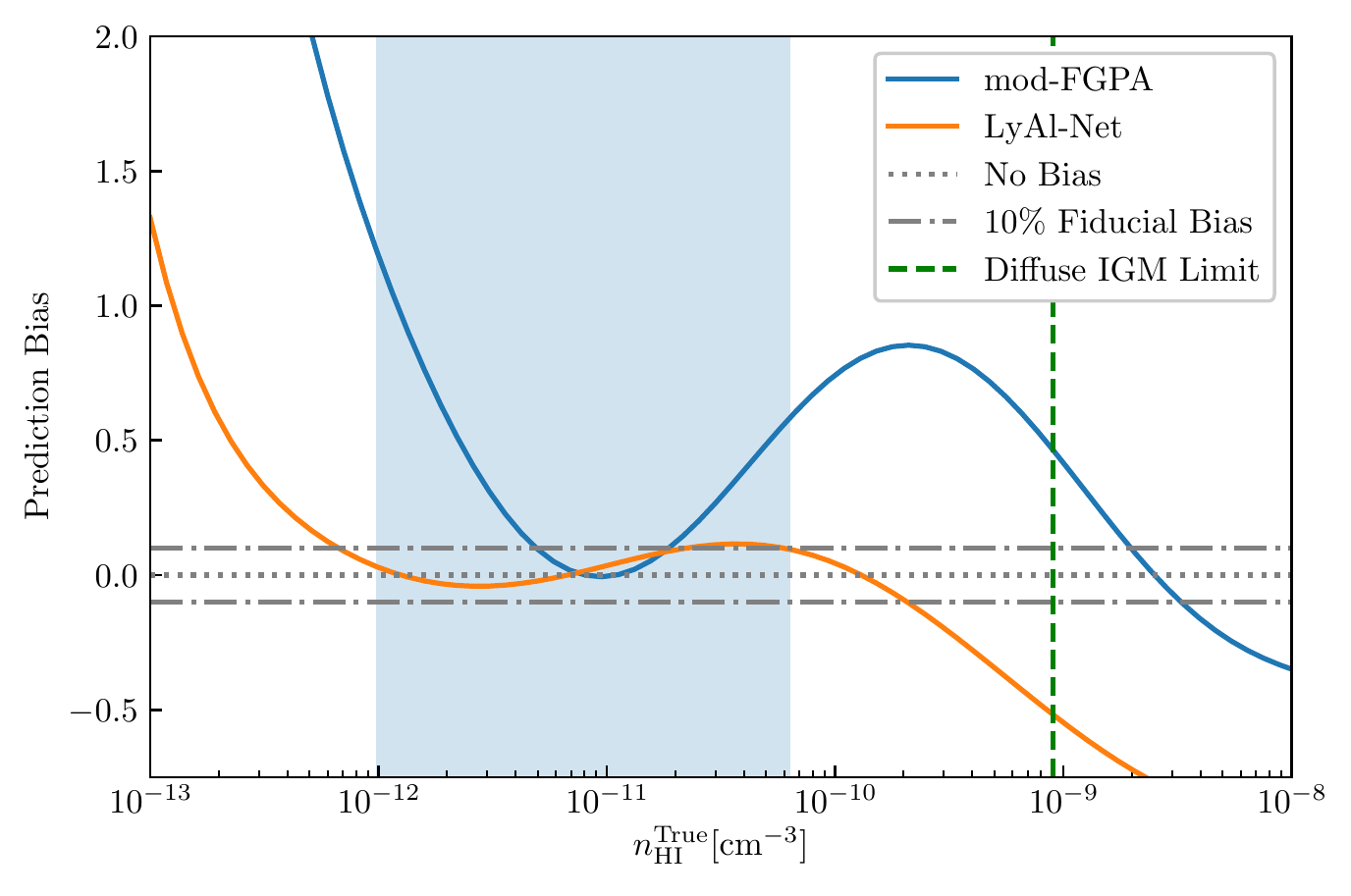}
     \caption{The emulated \nHI bias as a function of the true \nHI of \HnoAGN at $z=2.44$ of \LyNet (an orange solid line) and the modified-FGPA (a blue solid line). The bias of modified-FGPA 
     is only unbiased at \nHI  $\approx 10^{-11}\, \text{cm}^{-3}$.
     The \LyNet's bias shows a better prediction overall. It is mostly stable and stays inside the 10 \% fiducial bias levels for density range between 1st and 90th percentile, 
     highlighted in a light blue strip which corresponds to $1 \times 10^{-12}$ and $6 \times 10^{-11} \,\text{cm}^{-3}$ respectively. 
     The drastic behaviour of the bias outside the blue strip is expected since the extremities of the density on both sides rarely occur.
     }
         \label{fig:bias_nHI}
\end{figure}

We now follow the model evaluation procedure in Section~\ref{sec:quality} to compute the true number density's joint and conditional probability distribution from \HnoAGN and the emulated number density. The estimated probability distribution function $P(\log_{10} n_{\text{HI}}^\text{Pred}|\log_{10} n_{\text{HI}}^\text{True})$ is shown in Figure \ref{fig:posterior_nHI} with contour lines representing the different levels of probability. 
The green dashed line, referred to as $\mu \left(\log_{10}{ n_{\text{HI}}}^\text{Pred} | n_{\text{HI}}^\text{True} \right)$ is the mean predicted neutral hydrogen density from the conditional probability, along with the 68\% confidence interval represented in the green band. This is modelled after the fact that the conditional probability is quite close to a Gaussian distribution. 
The red dashed line is given as a reference where $n_{\text{HI}}^\text{Pred} = n_{\text{HI}}^\text{True}$. We further added a red band representing the upper and lower boundary of 10\% error from the mid-point as a qualitative reference to detect the mean bias of the emulator, which we call the prediction bias. The contour levels and the mean imply that the prediction behaves well within the main region of the dataset, given that the 0th and 99th percentiles are $4.5 \times 10^{-9}$ and $3.8 \times 10^{-13} \,\text{cm}^{-3}$ respectively. 
The bias only starts to occur when $n_{\text{HI}} < 10^{-12}\,\text{cm}^{-3}$ and $n_\text{HI} > 3\times10^{-9}\,\text{cm}^{-3}$ which might be caused by the longer distribution tail and sample variance affecting the KDE itself. 
To get a better picture of the prediction bias, we define the bias as a function of true value as 
\begin{equation}
   \mathrm{Bias} = 
   \left[ 
10^{\mu \left(\log_{10}{{ X_{\text{HI}}^\text{Pred} | \log_{10}{X_{\text{HI}}^\text{True}}}} \right)} - X_\text{HI}^\text{True}
   \right] \frac{1}{X_{\text{HI}}^\text{True}},
\end{equation}
where $X_\text{HI}$ is a gas parameter.
We visualise this function for the emulated \nHI in Figure~\ref{fig:bias_nHI}, and it shows the prediction bias where $ 10^{-12} \,\text{cm}^{-3}< \nHI < 10^{-10} \,\text{cm}^{-3}$  stay within $10\%$ fiducial bias. By comparing \LyNet to the benchmark, the prediction from modified-FGPA is unbiased only around the density \nHI  $\approx 10^{-11}\, \text{cm}^{-3}$, and fail to capture the broader range of the density.

While one-point statistics show that \LyNet performs well, we extend our analysis to the two-point statistics of the emulated fields. Before calculating the power spectra, we mask the extreme values for emulated and true \nHI to zero. We based this on the fact that the bias is starting to occur in the tails of the distribution due to the fewer data points. Furthermore, masking has to be done because power spectra are very sensitive to extreme values (see Figure~\ref{fig:pk_nHI_comparisons}). Therefore, we implement three different masking ranges and refer to them as \textit{fidelity range}: \textsc{low}, \textsc{medium}, and \textsc{high}. Table~\ref{tab:Pk_masking_ranges} summarises the threshold values used in this process. These masking values are arbitrary, so we can identify the regions in which \LyNet performs best. We emphasise that the maximum number of masked voxels is less than $1\%$. Therefore, this should not artificially increase the correlation rate $r(k)$.
\begin{table}[]
\caption{The different density ranges used in the masking process to evaluate the \LyNet performance where the upper-density boundaries are based on 99.95th, 99.9th,and 99.5th percentile for \textsc{low},\textsc{medium},and \textsc{high} fidelity range respectively. The lower boundary of neutral hydrogen density is $4.36 \times 10^{-15} \mathrm{cm}^{-3}$ and for temperature is 600 K. The maximum number of masked voxels is less than $0.5\%$, therefore, this should not artificially increase the cross-correlation $r(k)$.} 
\centering
\begin{tabular}{@{}ccc@{}}
\toprule
\multirow{2}{*}{\textbf{Fidelity Range}} & \multicolumn{2}{c}{\textbf{Upper Limit Masking Value}}                      \\ \cmidrule(l){2-3} 
                                         & \textbf{Density $[\mathrm{cm}^{-3}]$} & \textbf{Temperature $[\mathrm{K}]$} \\ \midrule
\textsc{low}                             & $5.16 \times 10^{-7}$                 & $3.08 \times 10^{6}$                \\
\textsc{medium}                          & $1.79 \times 10^{-8}$                 & $5.92 \times 10^{5}$                 \\
\textsc{high}                            & $1.10 \times 10^{-9}$                 & $8.03 \times 10^{4}$                  \\ \bottomrule
\end{tabular}
%move this up, please
\label{tab:Pk_masking_ranges}
\end{table}

\begin{figure}
   \centering
    \includegraphics[width=\hsize]{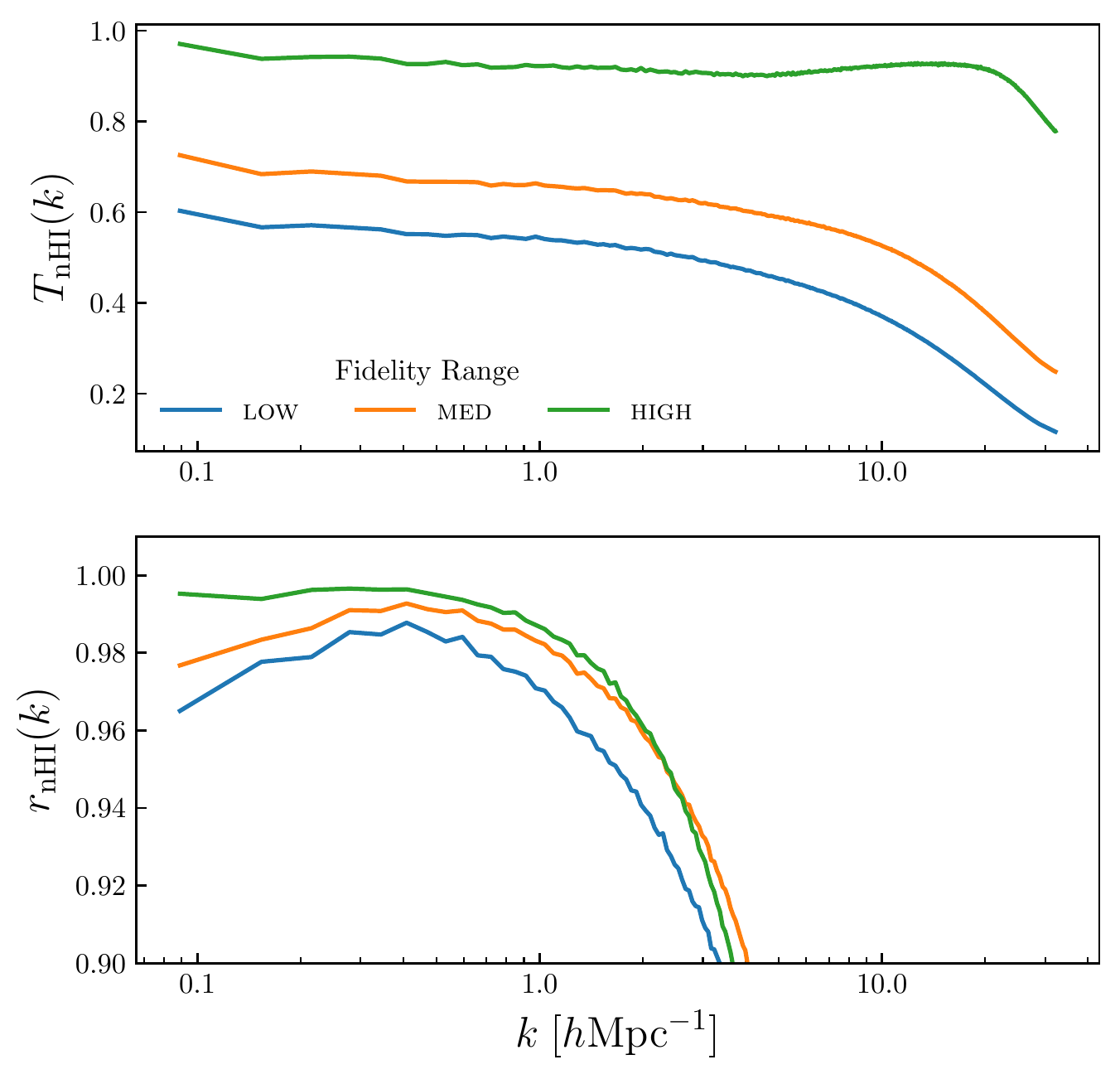}
    \caption{\HnoAGN prediction quality assessment using two-point correlations for a different choice of fidelity range for $\nHI$. \textbf{Top panel:} We show the transfer functions $T_\nHI(k)$ for the different for \textsc{low} (blue solid line), \textsc{med} (orange solid line), and \textsc{high} (green solid line). The descriptions are given in Table~\ref{tab:Pk_masking_ranges}. \textbf{Bottom panel:} We show the correlation rate $r_\nHI(k)$ for the same fidelity ranges as in the top panel, with the same colour convention. }
    \label{fig:masked_2point_nHI}
\end{figure}

We compare the transfer function and cross-correlation using different fidelity ranges illustrated in Figure \ref{fig:masked_2point_nHI}. The transfer function is approximately stable for all fidelity ranges while $T_\mathrm{HI}(k)$ of \textsc{high} fidelity performs the best and stays around 0.9 up to $k \approx 11\hMpc$. The cross-correlation functions $r_\mathrm{nHI}$ show the convergence of all fidelity ranges with the amplitude staying above $0.9$ up to $k \approx 3\hMpc$. Based on two-point statistics, this assessment shows that the best performance of the \LyNet is up to $\nHI \leq 1.10 \times 10^{-9}$ $\mathrm{cm}^3$, which is around the estimated diffuse IGM limit. 
\footnote{We estimated the diffuse IGM limit of neutral hydrogen density by scaling the \cite{Martizzi_2019} $ n_\mathrm{H}$ limit; $n_\mathrm{HI} = \alpha n_\mathrm{H}  = 10^{-4}(1+z)\;\text{cm}^{-3}$  where $\alpha = \text{mean}(n_\text{HI}/n_\text{H})$, obtained directly from \HnoAGN simulation at $z=2.43$.} 

\subsection{HI gas temperature (\TempHI)}
\label{sec:result-temp}

We perform the same analysis for the HI gas temperature as the density. We show the bottom panel in Figure~\ref{fig:Pred_sample} a slice of through the temperature field in the \HnoAGN simulation, the emulated temperature from dark matter density and the logarithmic difference between those two fields. The emulated and true temperature also agree without any apparent defects caused by the \LyNet. The logarithmic difference map of temperature shows that the map is, on average better than density, with a lower relative error for dense regions.

\begin{figure}
   \centering
   \includegraphics[width=\hsize]{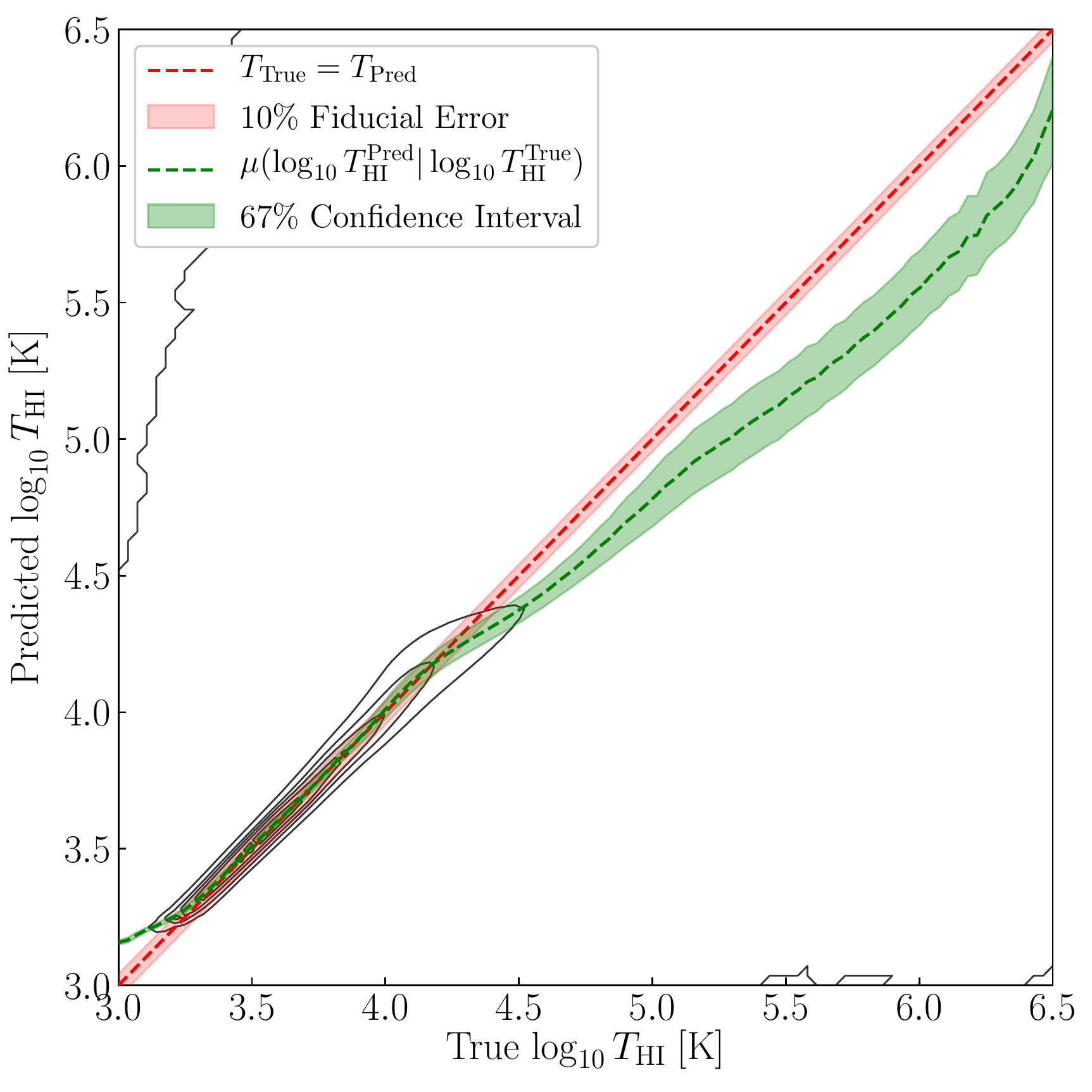}
      \caption{The conditional probability of the predicted decimal logarithm of temperature given the true temperature of the \HnoAGN where the contour lines represent different probabilities. The diagonal red-dashed line indicates the unbiased relation between those two temperatures alongside a fiducial 10\% error budget in shaded red. The dashed green line indicates the mean from the conditional distribution alongside the 67\% confidence interval (in shaded green).}
         \label{fig:posterior_temp}
\end{figure}

\begin{figure}
   \includegraphics[width=\hsize]{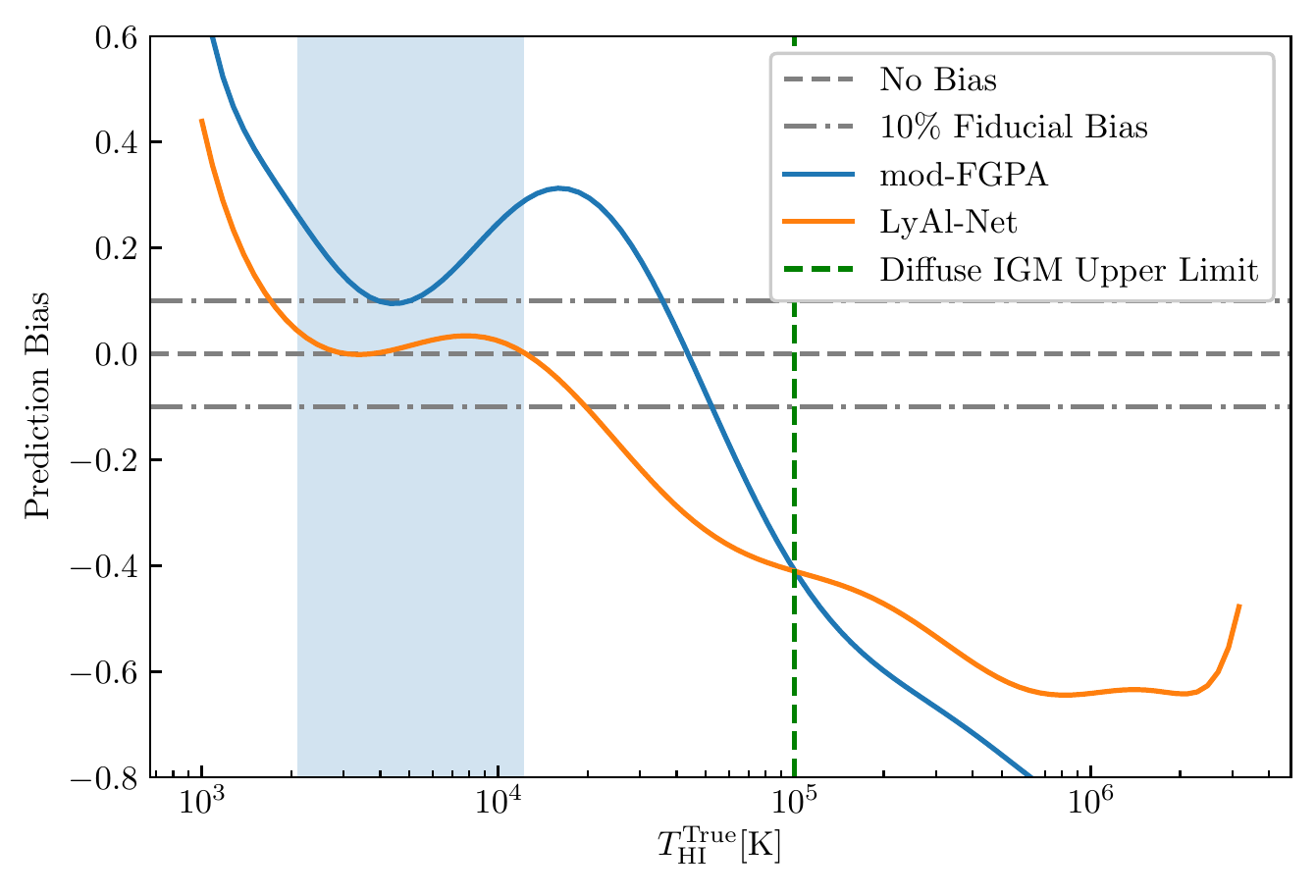}
      \caption{The emulated \TempHI bias as a function of the true \TempHI of \HnoAGN at $z=2.44$ of \LyNet (in an orange solid line) and the modified-FGPA (a blue solid line). 
      The modified-FGPA bias lies outside the 10 \% fiducial bias levels for all temperature range.
      While the bias of \LyNet is mostly stable and stays inside the 10 \% fiducial bias levels for temperature values between 1st and 90th percentile highlighted in a light blue strip which corresponds to $2\times 10^3$~K to $2\times 10^4$~K. 
      }
         \label{fig:bias_Temp}
\end{figure}

As for the density section, we also compute the emulated temperature's joint and conditional probability distribution given the true temperature, $P(\log_{10} T_{\mathrm{HI}}^\text{Pred}|\log_{10} T_{\mathrm{HI}}^\text{True})$ illustrated in Figure \ref{fig:posterior_temp}, and we follow the same colour scheme as \nHI. 
The contour lines represent different probability levels, the green dashed line is the mean predicted temperature, and the green band is the 68\% confidence interval taken from its maximum. The red dashed line is given as a reference where $T_{\mathrm{HI}}^\text{Pred} = T_{\mathrm{HI}}^\text{True}$ with a red band representing the 10\% fiducial error from the midpoint as a qualitative reference. Visually, the mean values of the emulated temperature appear to have a small bias across the temperature range, which becomes more prominent in the temperature regime. The details of the emulated temperature bias as a function of the true values are illustrated in Figure~\ref{fig:bias_Temp} in a solid orange line. Thus, we consider our emulator to be nearly unbiased and stable in the range of $2\times10^3 \;\text{K}<\TempHI<10^4 \;\text{K}$. The overprediction only appears when for the extreme temperature above $10^4$ K. In addition, \LyNet performs impressively better than modified-FGPA, which suffers from prediction bias in all ranges of the temperature value.


\begin{figure}
\includegraphics[width=\hsize]{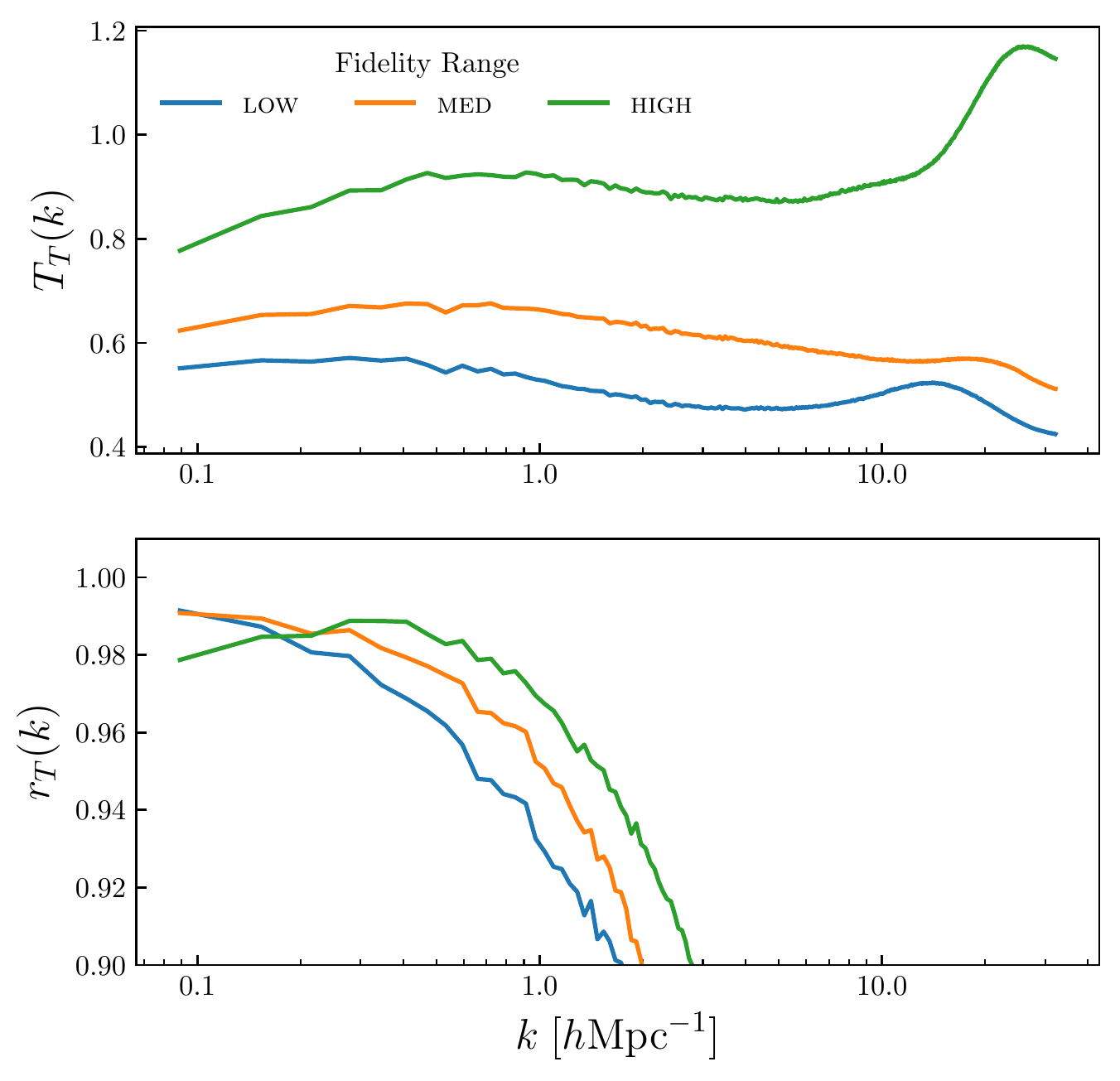}
    \caption{\HnoAGN prediction quality assessment using two-point correlations for different choices of fidelity range for emulated \TempHI. \textbf{Top Panel:} The transfer functions comparison for \textsc{low}, \textsc{med}, and \textsc{high} fidelity range in blue, orange and green respectively. The temperature range is indicated in Table~\ref{tab:Pk_masking_ranges}. \textbf{Bottom Panel:} A comparison of correlation rates of the same fidelity ranges with the same colour scheme.}
        \label{fig:masked_Tk_Temp}
\end{figure}

To estimate the fidelity range of \LyNet, we also picked three masking ranges, summarised in Table~\ref{tab:Pk_masking_ranges}. From Figure~\ref{fig:masked_Tk_Temp}, the transfer function indicates the amplitude 
stays well greater than $0.90$ up to $k \approx 10$ \hMpc  for the \textsc{high} fidelity ranges, with a small drop at $k < 0.3$\hMpc. The spatial information measured by the cross-correlation function indicates $r > 0.9$ up to $k=2$\hMpc for the \textsc{low} and \textsc{med} fidelity range and up to $k=3$\hMpc in the \textsc{high} fidelity range. 
%--------------------------------------------------------------------
This implies the highest performance of the \LyNet lies in the scale of 14\Mpch to 63\Mpch scale for the temperature range $ 1.92 \times 10^3	< T_\text{HI} <  8.03 \times 10^4$ K.
%------------------------------------------------------
It is worth pointing out that the mean prediction bias is stable when the values are within the 1st and 90th percentile on both \nHI and \TempHI. This range is empirical evidence that \LyNet performs best in the main distribution of both fields, while the high bias only occurs in the rare and extreme environment in the tails of the distributions. Thus, this implies an intrinsic limitation of the neural network weighing the temperature and density in the tails of distribution less due to the lack of data points. Fortunately, in practice, the \Lya absorption is not sensitive to the highly dense region because the saturation occurs at a density lower than where the discrepancy happens (see Equation~\ref{eq:norm_flux}).

\subsection{The line-of-sight velocity of the HI gas ($v_{z,\mathrm{HI}}$)}
\label{sec:result-vlos}
\begin{figure}
   \centering
   \includegraphics[width=\hsize]{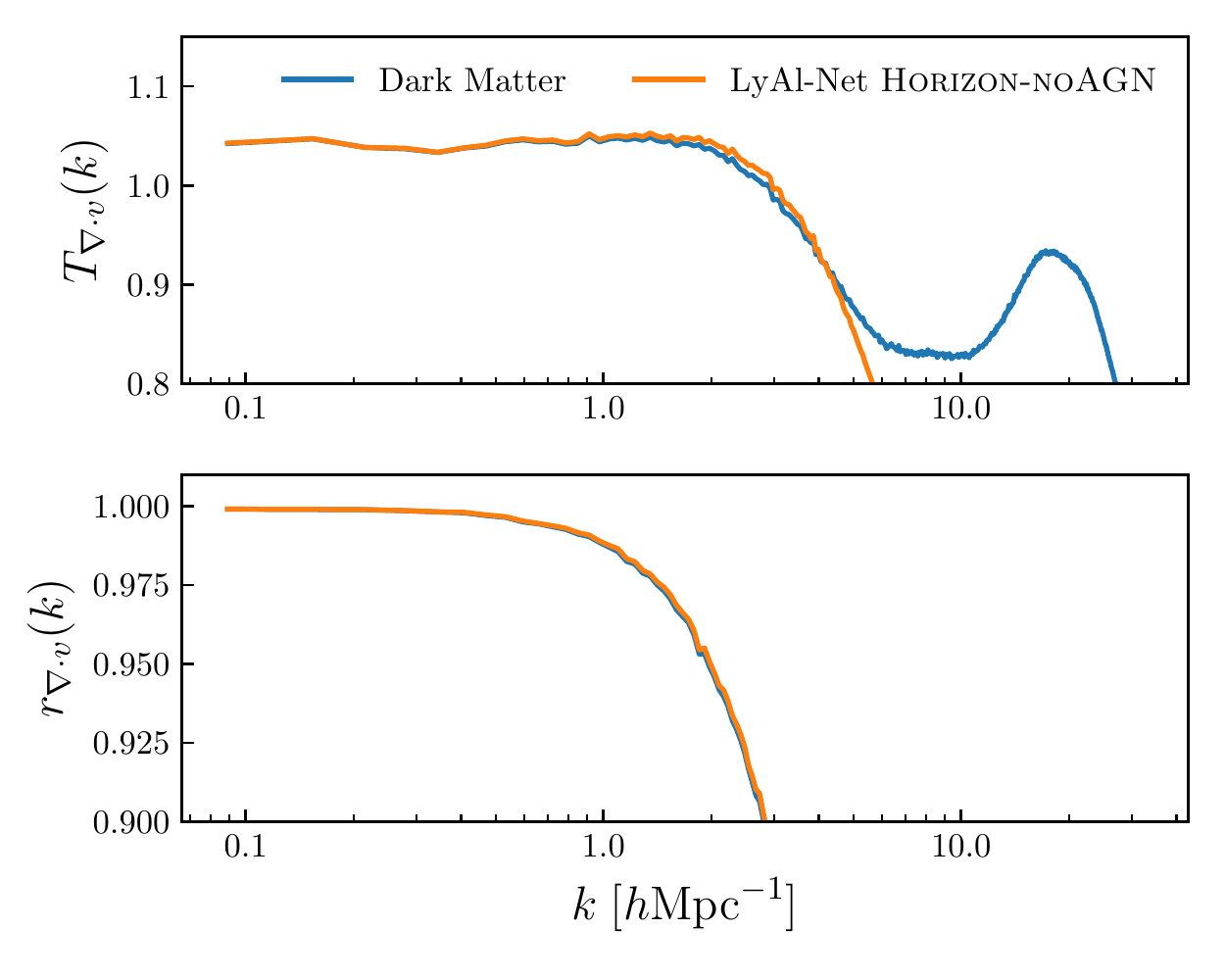}
   \caption{\HnoAGN prediction quality assessment using two-point correlations of velocity divergence from \HnoAGN dark matter velocity and emulated \HnoAGN HI velocity, labelled as \textbf{Dark Matter} in blue and \textbf{\LyNet} in orange respectively. \textbf{Top Panel:} Transfer functions comparison shows both have a similar amplitude. However, \HnoAGN dark matter velocity divergence outperforms where $k \gtrapprox 6 \hMpc$. \textbf{Bottom Panel:} Cross-correlation functions comparison shows both are identical, which implies that \LyNet does not improve the spatial distribution.
   }
    \label{fig:Pk_Divergence}
\end{figure}
The gas line-of-sight velocity, $v_{z,\mathrm{HI}}$, plays a significant role in the \Lya absorption by shifting the frequency of an incoming photon due to the Doppler effect (for a non-relativistic case). With inaccurate values of the velocity field, the emulated flux will result in a change in absorption rate and a shift of \Lya absorption in some frequencies. 
To estimate $v_{z,\mathrm{HI}}$, we can use the line-of-sight velocity of dark matter $v_{z,\mathrm{DM}}$ directly, because their velocity fields are nearly identical since the gravity from dark matter influences the IGM. This estimation is effective on large scales, but it is inevitable for the IGM velocity to decouple at small scales due to various shocks and feedback \citep{Pando_2004}. Therefore, naively using $v_{z,\mathrm{DM}}$ could potentially impact the accuracy of absorption at small scales, which means there is room for improvements to emulate $V_\mathrm{HI}$.
To show the decoupling effect empirically, we compare velocity divergences of $V_\mathrm{DM}$ to $V_\mathrm{HI}$ using the transfer function and cross-correlation in Figure~\ref{fig:Pk_Divergence} in a solid blue line. The transfer function $T_{\mathbf{\nabla} \cdot v}(k)$ of dark matter shows a steady amplitude $\approx 1.1$ where $k<2 \hMpc$, which means the dark matter velocity fluctuation is higher than the gas. The cross-correlation rate $r_{\mathbf{\nabla} \cdot v}(k)$ on the bottom panel shows the value drops below $0.95$ when $k>2 \hMpc$, meaning the IGM velocity starts to decouple with below 3 \Mpch.
Based on the idea that \LyNet should be able to correct the small-scale effect, we train \LyNet to predict a HI line-of-sight velocity using dark matter line-of-sight velocity as an input. 
We check whether \LyNet improves the results instead of simply adding more systematic effects to the dark matter velocity field and illustrate in a solid orange line in Figure~\ref{fig:Pk_Divergence}. 
Both methods show similar performance in the spatial distribution of cross-correlation of velocity divergences. However, the transfer function of dark matter shows better reliability in the high-$k$ regime, while \LyNet introduces a loss of amplitude over the same range. Despite this, we use $v_{z,\mathrm{DM}}$ as a proxy for $v_{z,\mathrm{HI}}$ and postpone the improvement for velocity prediction in future work.

\subsection{The effect from an absence of the baryonic feedback}
\label{subsubsec:Horizon-DM}

\begin{figure}
    \includegraphics[width=\hsize]{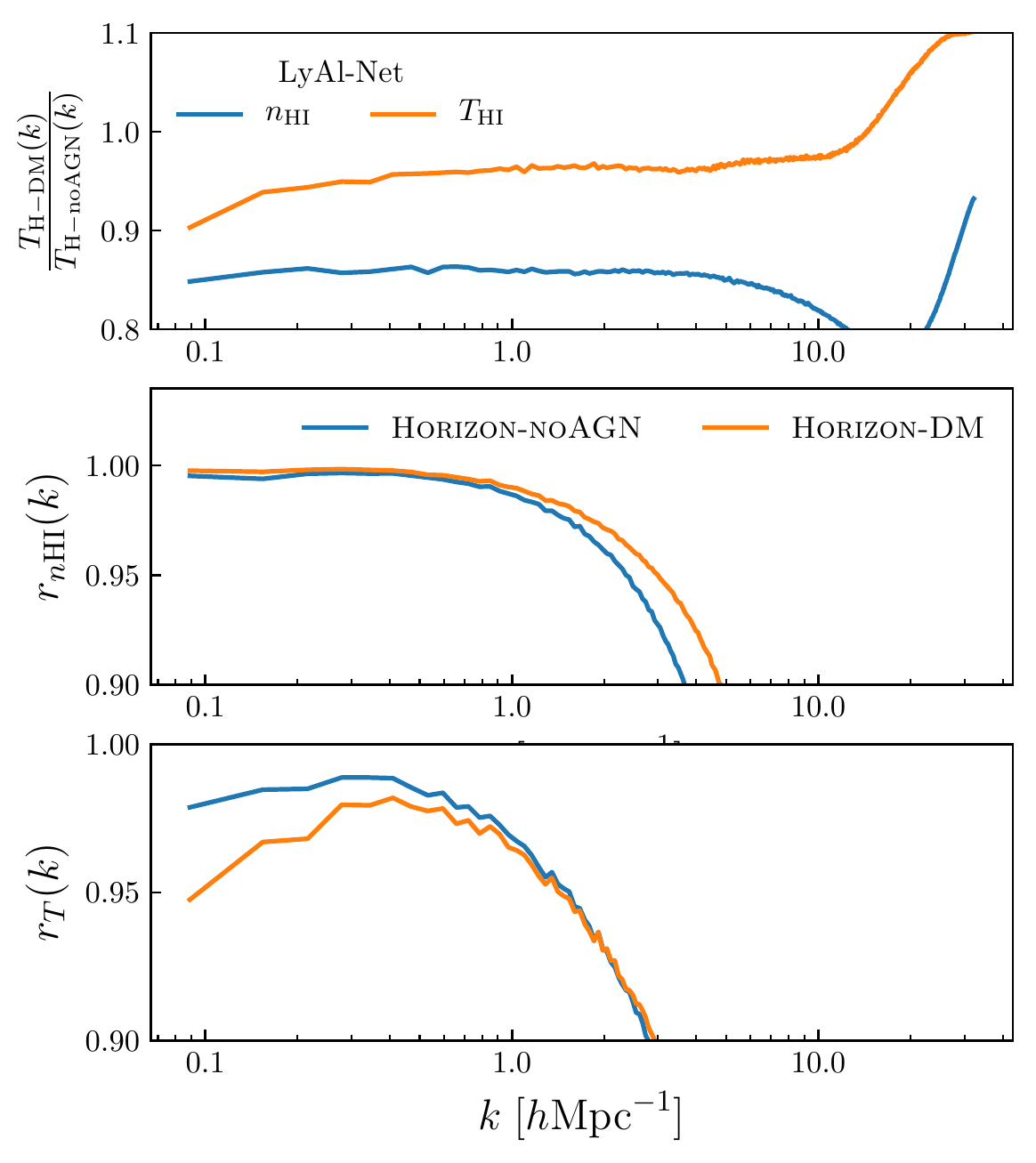}
    \caption{Impact of the alteration of dark matter fields from baryonic physics on the quality of the emulation. \textbf{Top Panel} shows the ratio of the transfer function of emulated fields using \HnoAGN and \HDM showing that the temperature does not suffer from the absence of baryonic feedback. However, the density of neutral hydrogen loses power, meaning the density is smoother. \textbf{Middle Panel} and \textbf{Bottom Panel} show comparisons of high-fidelity range cross-correlation functions of emulated density and temperature
    from \HnoAGN (illustrated previously in Figure~
    \ref{fig:masked_2point_nHI} and \ref{fig:masked_Tk_Temp}) and \HDM.}
    \label{fig:T-r_pureDM}
\end{figure}
With the additional complexity of the equations and the large increase in the number of timesteps, cosmological hydrodynamic simulations are largely more computationally expensive by at least an order of magnitude \citep{Peirani2017} than an $N$-body simulation. While it is cheaper to produce dark matter-only fields, neglecting the baryonic feedback can reduce the accuracy of the emulated gas fields on smaller scales. 
Since the main goal is to utilise \LyNet on pure dark matter $N$-body simulations for time efficiency, in this section, we explore how a pure dark matter $N$-body simulation can affect the quality of emulated fields. We then look at the impact of this approximation on the \Lya forest absorption in Section~\ref{subsec:Flux}.
For this test, we emulate HI density and temperature from \HDM. This simulation is a sibling simulation of \HAGN, which considers only pure dark matter dynamics. We use the same snapshot, i.e. the same redshift range, as the one used for the training of \LyNet. 
Since the neural network model is trained on \HnoAGN with the dark matter density transformation described by Equation~\eqref{eq:log_dm}, it allows \LyNet to be immune to the difference in dark matter particle mass of pure $N$-body and a full hydrodynamical simulation. We can use these emulated fields as benchmarks for the case of emulated fields from \HDM directly.

Let us introduce a new metric, which is the ratio of transfer functions,
\begin{equation}
    \frac{T_{x,\mathrm{H-DM}}(k)}{T_{x,\mathrm{H-noAGN}}(k)} = \sqrt{\frac{P_{x,\mathrm{H-DM}}(k)}{P_{x,\mathrm{H-noAGN}}(k)}},
    \label{eq:Transfer_fraction}
\end{equation}
where the a subscripts $x,\mathrm{H-noAGN}$ and $x,\mathrm{H-DM}$ are denoted as a gas field $x$ emulated using \HnoAGN and \HDM respectively.
We also use a \textsc{high} fidelity masking range introduced in Table~\ref{tab:Pk_masking_ranges} for both fields. 

The top panel of Figure~\ref{fig:T-r_pureDM} shows the ratio of the transfer functions for the emulated density and the temperature. 
The ratio is relatively stable across the whole range in the considered scales for the neutral hydrogen density. However, this stable value stays at $\approx 0.85$, which means the absence of baryonic feedback impacts the fluctuation of \nHI. The cross-correlation function of \nHI (middle panel) shows a slight increase in its accuracy at $k \approx 5\hMpc$. This is slightly better than the \HnoAGN and contradicts the expectation that the emulated fields from \HDM should be worse due to the lack of feedbacks.
On the other hand, the emulated temperature from \HDM performs much closer to using \HnoAGN because the ratio stays relatively stable at $0.95$. Its cross-correlation function shows that using \HDM slightly impacts the spatial distribution of the emulated field on a large scale. 
It is worth pointing out that when $k > 10\hMpc$, the transfer function ratio starts to deviate for both fields, which implies that the lack of baryonic feedback dominantly affects scales above 10\hMpc, i.e. 0.62\Mpch.
The accuracy reduction of emulated gas fields by using a dark matter simulation with different settings of baryonic effect is expected, especially when the shocks and feedbacks are absent. In Section~\ref{subsubsection:TNG-100:Transfer Learning}, we will discuss this in more detail and the possible treatment/improvements using IllustrisTNG where the difference of the simulation parameters is much more complicated.

\subsection{Lyman Alpha forest Absorption ($F$)}
\label{subsec:Flux}
In this section, we discuss the Lyman alpha forest prediction quality using emulated HI density and temperature using \HnoAGN and especially what happens when the baryonic feedback is absent using \HDM.

\begin{figure*}
   \centering
   \includegraphics[width=\hsize]{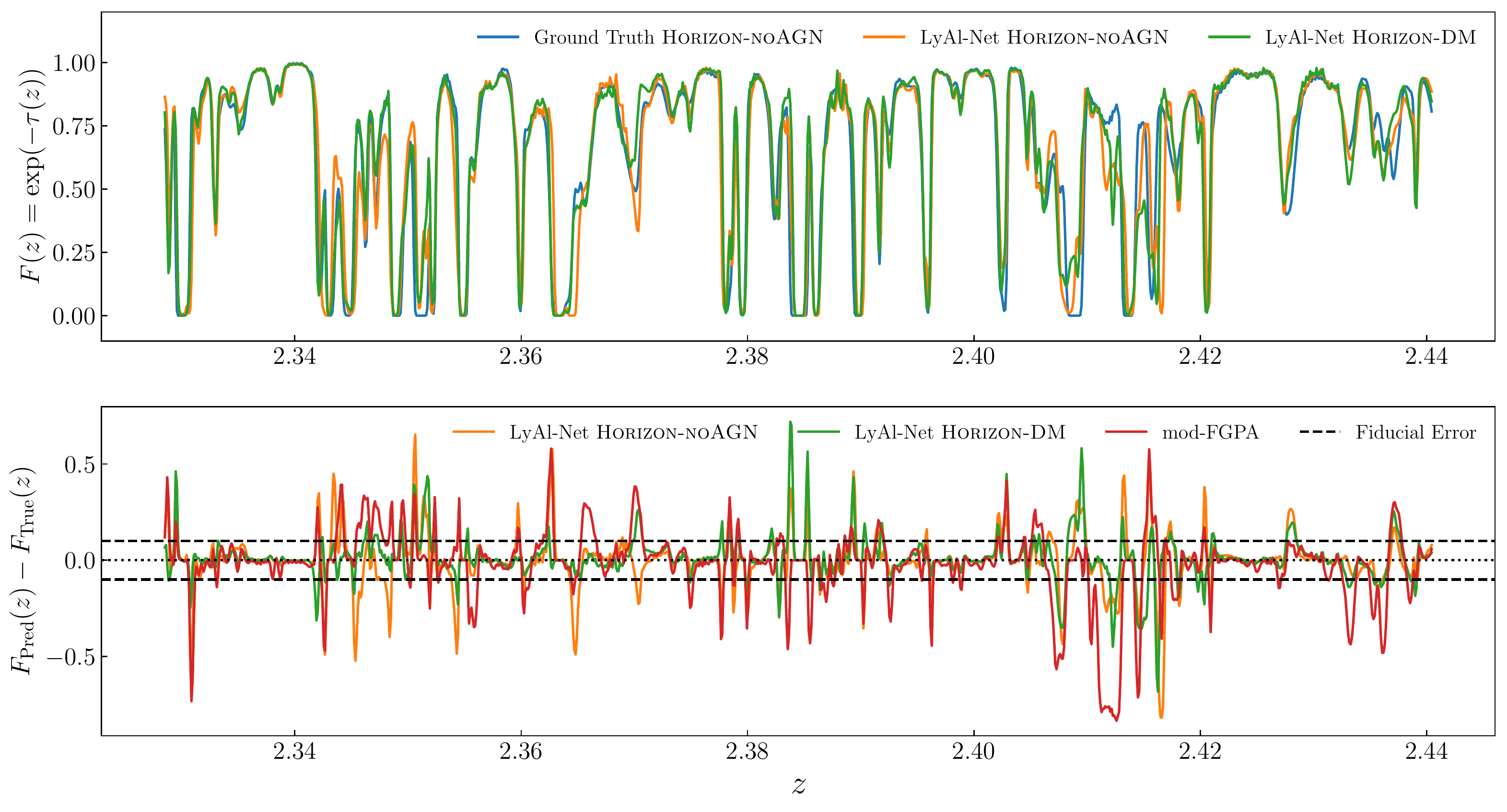}
      \caption{\textbf{Top Panel:}A comparison of \Lya forest normalised fluxes from the same sample skewer of the ground truth and emulated flux, labelled as Ground Truth and \LyNet, respectively. The fluxes are calculated from \HnoAGN and emulated HI parameters from \LyNet using \HnoAGN and \HDM dark matter. \textbf{Bottom Panel:} Residuals of emulated flux and true fluxes along with fiducial error plotted in dashed line at $-0.1$, and $0.1$.Both \LyNet and modified-FGPA show that the values mostly stay within the fiducial region while the residuals and peaks within the highly absorbed region with \LyNet being more stable.}
         \label{fig:sample_absorption}
\end{figure*}

\begin{figure}
    \includegraphics[width=\hsize]{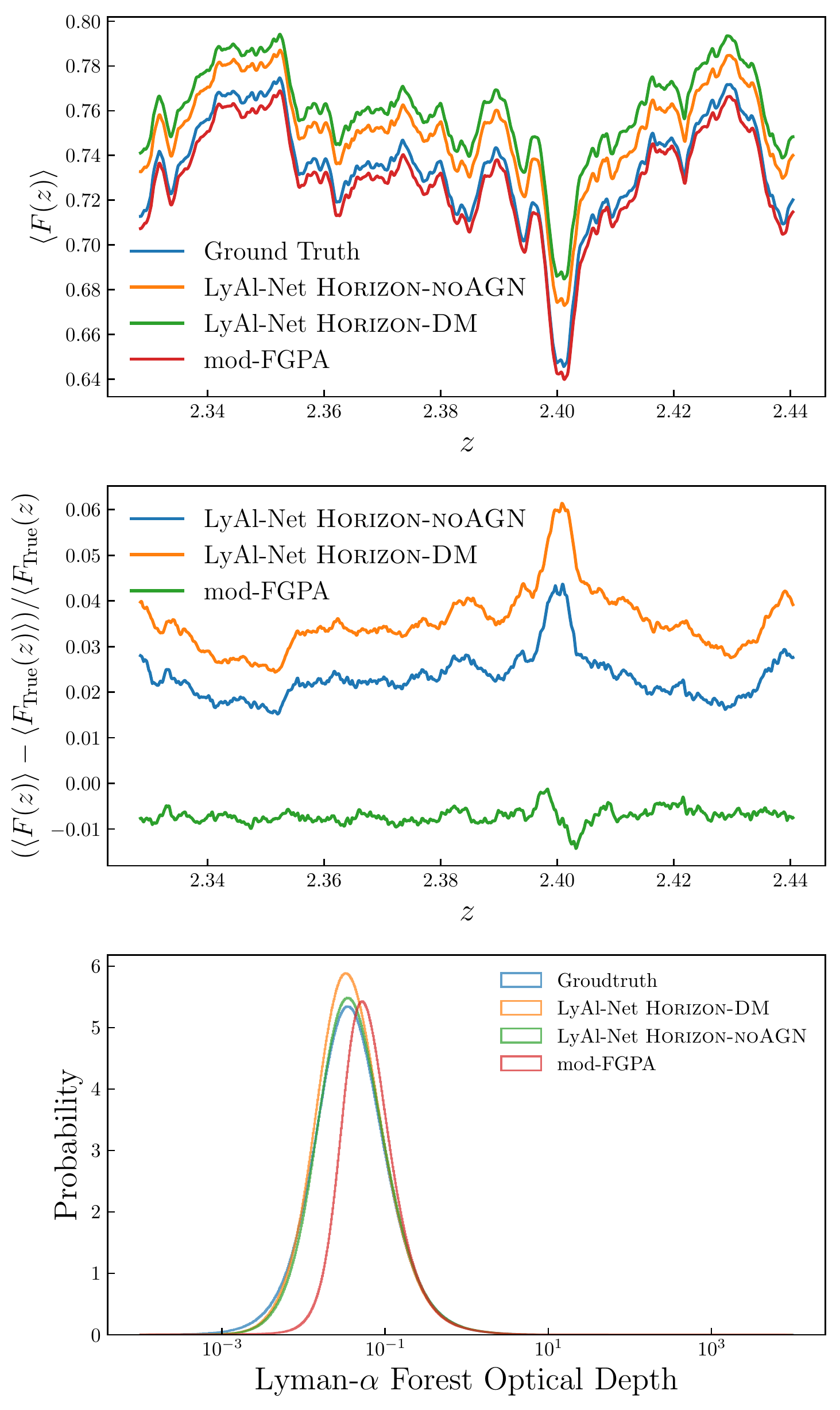}
    \caption{One-point statistics comparisons of \Lya absorption between \HnoAGN (Ground Truth) and emulated absorptions, using \LyNet and the benchmark model (mod-FGPA)
    \textbf{Top Panel}: A mean transmitted flux as a function of redshift, $\langle F(z) \rangle$, shows that \LyNet tracks well with the ground truth. 
    \textbf{Middle Panel}: A comparison of relative error of mean transmitted flux as a function of redshift between flux emulated. Both \LyNet's relative errors exhibit the same shape, while \HnoAGN has the mean closer to the ground truth. The benchmark has a closer mean transmitted flux. \textbf{Bottom Panel}: A comparison of optical depths 1D distribution  shows \LyNet flux using \HnoAGN and \HDM perform better than the benchmark model.
    }
    \label{fig:Flux_residual_horiz}
\end{figure}

We follow the Equation~\eqref{eq: opacity discre} and calculate the normalised \Lya absorption. Figure~\ref{fig:sample_absorption} shows \Lya absorption derived from the emulator for a sample skewer, and the flux computed from the gas parameters from \HnoAGN as a ground truth. The bottom panel shows the residual, defined as
\begin{equation}
    \text{Residual} \equiv \exp(-\tau_\text{True}) - \exp(-\tau_\text{Pred}) = F_\text{True} - F_\text{Pred}.
\end{equation}
We also include the flux residual estimated by FGPA as a benchmark model\footnote{We include the Voigt profile described in Section~\ref{subsec:FGPA}}. A visual inspection shows that the emulated spectra agree with the overall residual staying within the $\pm 0.1$ fiducial error. Some prominent errors exist in the highly absorbed region where \LyNet shows limited performance in extreme-density environments. However, it shows the overall prediction performance better than the FGPA model.
Nonetheless, Individual-skewer residuals do not provide enough information to assess the emulated field quality due to a few reasons. The absorption requires three different parameters which can affect different behaviours of the emulated flux:
\begin{enumerate}
     
    \item the density \nHI directly affects the amount of absorption;
    \item the temperature \TempHI affects the absorption width due to thermal broadening;
    \item the line-of-sight velocity affects the absorption frequency, which will appear as the a shift on spectra

\end{enumerate}
Therefore, to suppress the small-scale fluctuation and detect the systematic error of the emulation, we perform an ensemble average of the transmitted flux along the XY-plane to obtain the mean transmitted flux as a function of redshift ($\langle F(z) \rangle$) and illustrate in Figure~\ref{fig:Flux_residual_horiz}. The top panel of the figure shows that the fluctuations of the \LyNet flux versus the true flux (referred to as "ground truth") agree on all the redshift ranges. The only main difference is the amount of absorption from emulated fields. The bottom panel shows that the relative error of \LyNet using \HnoAGN and \HDM are approximately $2.5\%$ and $3.5\%$, respectively. This is directly affected by the underprediction affecting the column density in the high matter density region of \LyNet. The relative error also shows the prominent fluctuations around redshift $z \approx 2.35$, $2.40$, and $2.43$. The stochastic components can explain this fluctuation from the various feedbacks that \LyNet might ignore. To alleviate these effects would require designing a new loss function that would mitigate the impact of large deviations during the training. The relative error of modified-FGPA is approximately $1\%$. However, the 1D distribution  of \Lya optical depth $\tau$ shows that \LyNet outperforms the benchmark model, particularly in the low optical depth region where the modified-FGPA fails to predict.

\begin{figure}
    \includegraphics[width=\hsize]{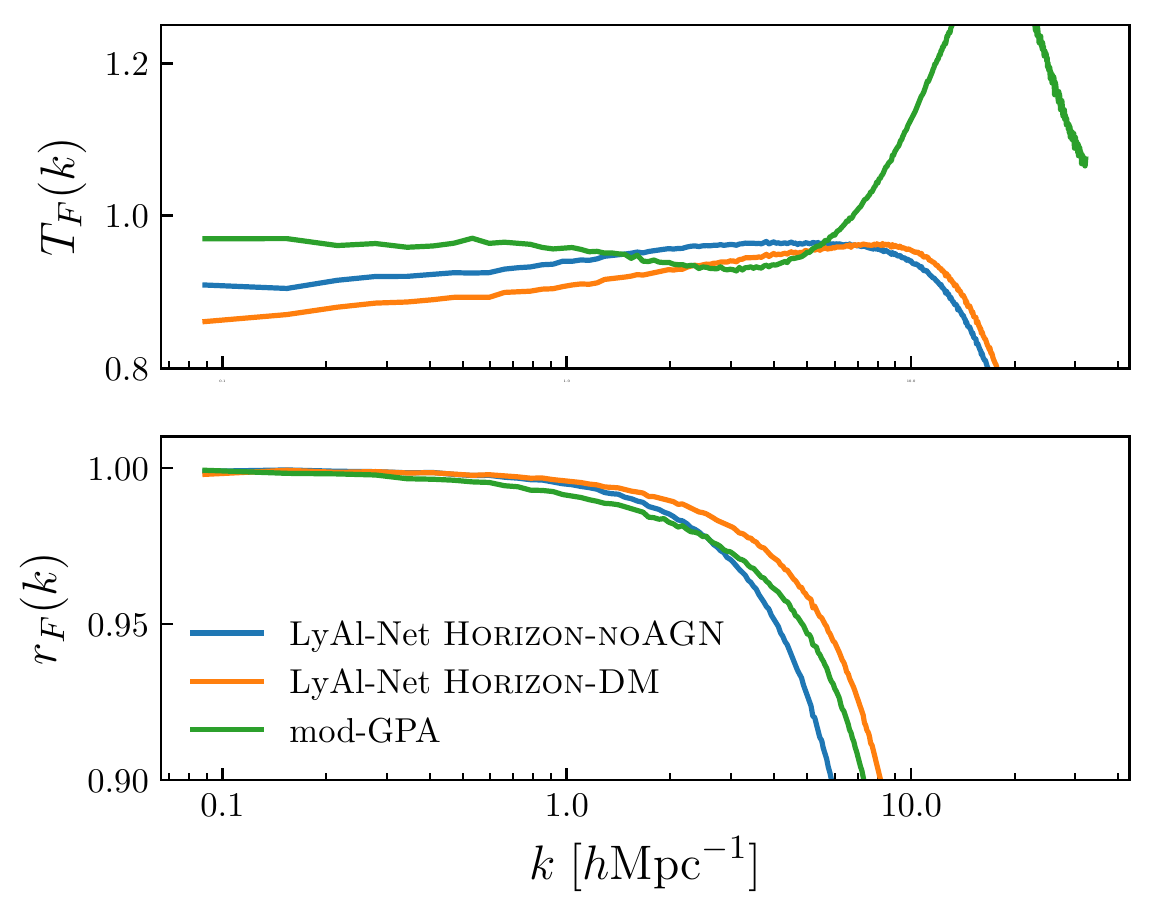}
    \caption{The comparisons of transfer functions of and cross-correlation function (Top panel) and cross-correlation function (Bottom panel) of the  emulated using \LyNet on \HnoAGN, \textsc{Horizon-DM}, and modified-FGPA. The transfer function of modified-FGPA noticeably diverges when $k>3$ which implies that \LyNet performs better and more stable in the smaller scales.} 
    \label{fig:Pk_flux_Horiz}
\end{figure}

For the two-point summary statistics illustrated in Figure~\ref{fig:Pk_flux_Horiz}, the transfer function using \HnoAGN (in blue) shows that the amplitude stays above $0.90$ up to $k \approx 10\hMpc$. We note that $T_F(k)$ the initial drop followed by the slow rise in large scales ($k \approx 0.1 \hMpc$) implies the predicted flux is smoother than the ground truth. When turning baryonic feedback off, the flux emulated from \HDM behaves the same way as \HnoAGN with a smaller amplitude. This reduction in the amplitude is the direct impact of the \nHI underprediction discussed in Section~\ref{subsubsec:Horizon-DM}. Meanwhile, the modified-FGPA model shows a higher amplitude in larger scales which also reflects on the ensemble average in Figure~\ref{fig:Flux_residual_horiz} being closer to \HnoAGN. The transfer function, however, noticeably diverges when $k>3$, which implies that \LyNet successfully outperforms the benchmark model on smaller scales.
The cross-correlation function on the bottom panel shows a remarkable agreement between the emulated fluxes and the ground truth up to $k \approx 6\hMpc$ for using \HnoAGN and $k \approx 9\hMpc$ for \HDM. the modified-FGPA also shows a similar level of agreement as \LyNet. However, the spatial distribution is better with \HDM. We can think of two possible reasons for this: the cross-correlation of \HDM is slightly better (see Figure~\ref{fig:T-r_pureDM}), or its velocity field is coincidentally closer to the velocity field of the gas. We will investigate this in future work.

We now consider if the inaccuracies of the model would be problematic to model observations. 
For a practical example, \cite{Bautista_2015} modelled mock quasar spectra as
\begin{equation}
\label{eq:Bautista}
    f(\lambda)=\left\{[F(\lambda) \cdot C(\lambda)] * \tilde{W}\left(\lambda, R_p, R_w\right)+N(\lambda)\right\} \cdot M(\lambda)+\delta f_{\text {sky }}(\lambda) .
\end{equation}
Where $f(\lambda)$ is a quasar flux, $F(\lambda)$ the \Lya absorption, $C(\lambda)$ is a quasar continuum, $\tilde{W}\left(\lambda, R_p, R_w\right)$ is a BOSS pixelisation kernel, $N(\lambda)$ is a noise from \cite{Bautista_2015},  $M(\lambda)$ is a correction linear function, and $\delta f_{\text {sky }}(\lambda)$ is the added sky subtraction residual. 

Let us consider a \Lya absorption from full hydrodynamical simulation to be $F(z)$. the output from \LyNet can be expressed as
\begin{equation}
\label{eq:LyAl flux}
    F(z)_{\LyNet}  = F(z)B(z)
\end{equation}
where $B(z)$ is an arbitrary intrinsic bias from \LyNet responding to the $N$-body simulation input. Therefore, one has to take into account a systematic bias from the prediction model. Therefore, the Equation~\eqref{eq:Bautista}, $F(\lambda)$ now has an extra term of $B(\lambda)$, which has to be treated.

If we assume as a minimum requirement that the QSO's spectra signal-to-noise ratio be $\mathrm{SNR}=10$ and further take the bias to have a linear scaling with $2.5 \%$ bias. Hence, the requirement for SNR will be changed to $10.3$.

%--------------------------------------------------------------------
\section{Transfer Learning with IllustrisTNG}
\label{sec:transfer_learning}

In the previous sections, we test the performance of the direct application of \LyNet on simulations. The real universe, however, is not likely to follow any of those. Therefore, it is essential to check the portability of this model. To guarantee this, we need a framework that generalises to other physical feedback and other updates of the gas physics. 
In this section, we explore the efficiency of the porting of the trained \LyNet, which is trained on \HnoAGN, to different gas physics. We are using IllustrisTNG as a benchmark, and a validation set, to test the accuracy of the predicted density and temperature fields. We first test this by applying \LyNet straightaway (out of the box), and then we will discuss the possibility of improving the prediction results by using a technique inspired by transfer learning.

\subsection{TNG100 specification}
IllustrisTNG is a set of simulations with different settings and resolutions. We choose TNG100 since it has the volume closest to the \HnoAGN simulation. TNG100 is a hydrodynamic cosmological simulation where its volume spans a cube with a dimension of $75\Mpch$ on each side. The simulation relies on AREPO, which solves the dynamics equation on a moving unstructured mesh defined by the Voronoi tessellation of a set of discrete points  \citep{Volker2010}. 
We picked snapshot 28, which corresponds to the redshift $z=2.58$. It is the closest snapshot to the trained \LyNet with \HnoAGN $z=2.5$. 

\subsection{Precomputation}

As mentioned above, the data structure of TNG100 is slightly different from \HnoAGN, as it is based on the Voronoi tessellation. To derive the dataset of the hydrodynamic quantities for our test, we deploy the same technique using the same adaptive filter to efficiently assign the dark matter and gas fields into a mesh structure with the same resolution as the \HnoAGN training set ($\sim 9.78 \Mpch$) which yields $768 \times 768 \times 768$ voxels.

We derived the temperature of the gas from the internal energy, $u$, of the TNG100 simulation by using the relation  \footnote{We followed the guide from the official The IllustrisTNG Project (\url{https://www.tng-project.org/data/docs/faq/##gen5}).}
\begin{equation}
  T= \mu (\gamma-1)  \frac{u}{k_B}  \;,  
\end{equation}
where $T$ is the temperature of the gas in each cell, $\gamma = 5/3$ is the adiabatic index for an ideal monoatomic gas, $k_B$ is the Boltzmann constant, and $\mu$ is the mean molecular weight. The mean molecular weight is derived from the electron abundance, a ratio of the number density of free electrons and a total hydrogen number density \citep{Kim2022}: This relation is expressed by, 
\begin{equation}
    \label{eq:mean_mole_weight}
    \mu = \frac{4m_p}{1+3X_H+4X_Hx_e}\;,
\end{equation}
where $m_p$ is proton mass and $x_e = n_e/n_H$ is the electron abundance which is a fraction of free electrons with respect to the number density of total hydrogen, $X_H$ is the hydrogen mass fraction which we assume to be constant and uniform. 

The database of the IllustrisTNG simulations does not publicly provide the neutral hydrogen density. We instead choose to estimate it from the total hydrogen density, which is available. Therefore, we estimated the density of neutral hydrogen, $\nHI_\mathrm{mock}$, by scaling it from its mean relative abundance in the \HnoAGN simulation. This can be expressed as
\begin{equation}
  \nHI_\mathrm{mock}^\mathrm{TNG100} = \Bigg\langle \frac{\nHI^\mathrm{HnoAGN} }{ n_\mathrm{H}^\mathrm{HnoAGN} }\Bigg\rangle n_\mathrm{H}^\mathrm{TNG100}\;. %
  \label{eq:nH_density}
\end{equation} The above equation is an important simplification of the complicated physics of the hydrogen ionisation state. In future work, we intend to better model this effect for the training set through adequate computation of this equilibrium and matching it to \HnoAGN.
For a performance comparison, we train the \LyNet to predict the total hydrogen density, $n_H^\mathrm{HnoAGN}$, with the same \HnoAGN, allowing us to scale the density using Equation~\eqref{eq:nH_density} to obtain the TNG100 mock neutral hydrogen density.  

\subsection{Assessment of \LyNet capabilities}
\label{sec:assessment_LyAl}
We now discuss the performance of the emulated neutral hydrogen density and temperature using \LyNet on dark matter overdensity on TNG100.

\subsubsection{Hydrogen Density}

\begin{figure}
    \centering 
    \includegraphics[width=\hsize]{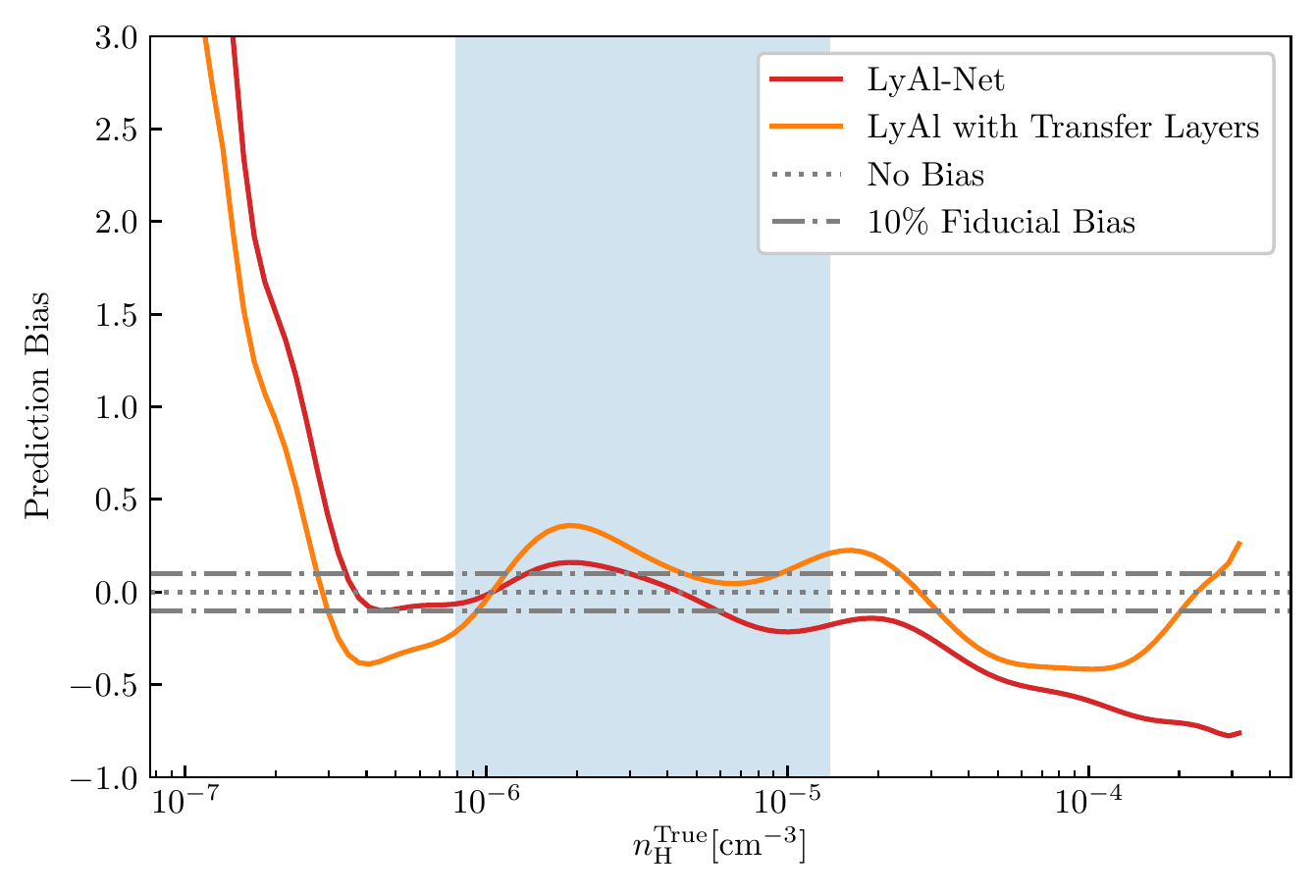}
    \caption{The comparisons of the prediction bias before and after applying the transfer learning on a TNG100 snapshot 28 (Sec \ref{subsubsection:TNG-100:Transfer Learning}).The light blue band represents the density within the 1st and 90th percentile. It shows the general improvement in the high-density regime after the 0th percentile while maintaining a similar bias across the density range compared to the out-of-the-box \LyNet.}
    \label{fig:Bias_nH_TNG100}
\end{figure}

\begin{figure}
    \includegraphics[width=\hsize]{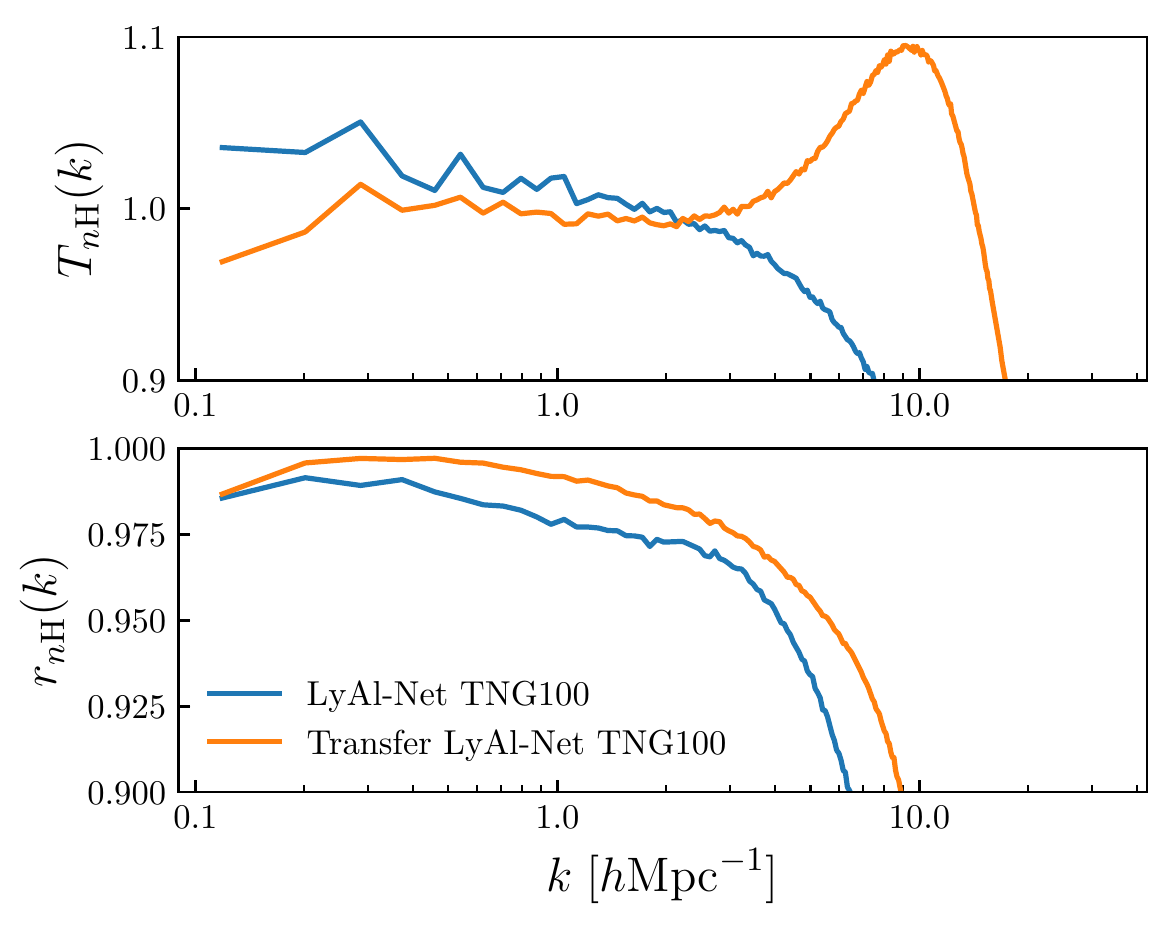}
    \caption{Top panel: transfer functions and Bottom panel: cross-correlation function of the neutral hydrogen temperature emulated using \LyNet. The blue line represents the performance of \LyNet out-of-the-box, and the orange line represents the performance of \LyNet with transfer layers. We also used the (high fidelity) 99.5th percentile range masking corresponds to $1.13\times10^{-7}$ to $1.26\times10^{-4}$ $\text{cm}^{-3}$}.
    \label{fig:pk_TNG_nH}
\end{figure}

We perform the same KDE analysis for one-point statistics on the 400 slices ($768 \times 768 \times 400$ voxels). Figure~\ref{fig:Bias_nH_TNG100}  shows that \LyNet's prediction bias remains relatively stable and stays around the $10\%$ fiducial bias in the first and ninetieth percentile, as highlighted in blue. The overall prediction is more prominent compared to \HnoAGN counterpart, and this is expected to occur. Figure~\ref{fig:pk_TNG_nH} illustrates a transfer function (top panel) and correlation rate (bottom panel), which stay above $90 \%$ up to $k \approx 7\hMpc$ and $6\hMpc $ 
respectively. In addition, the accuracy drops from $k \approx 10$ \hMpc compared to \HnoAGN. 

\begin{figure}[!bht]
    \centering
    \includegraphics[width=\hsize]{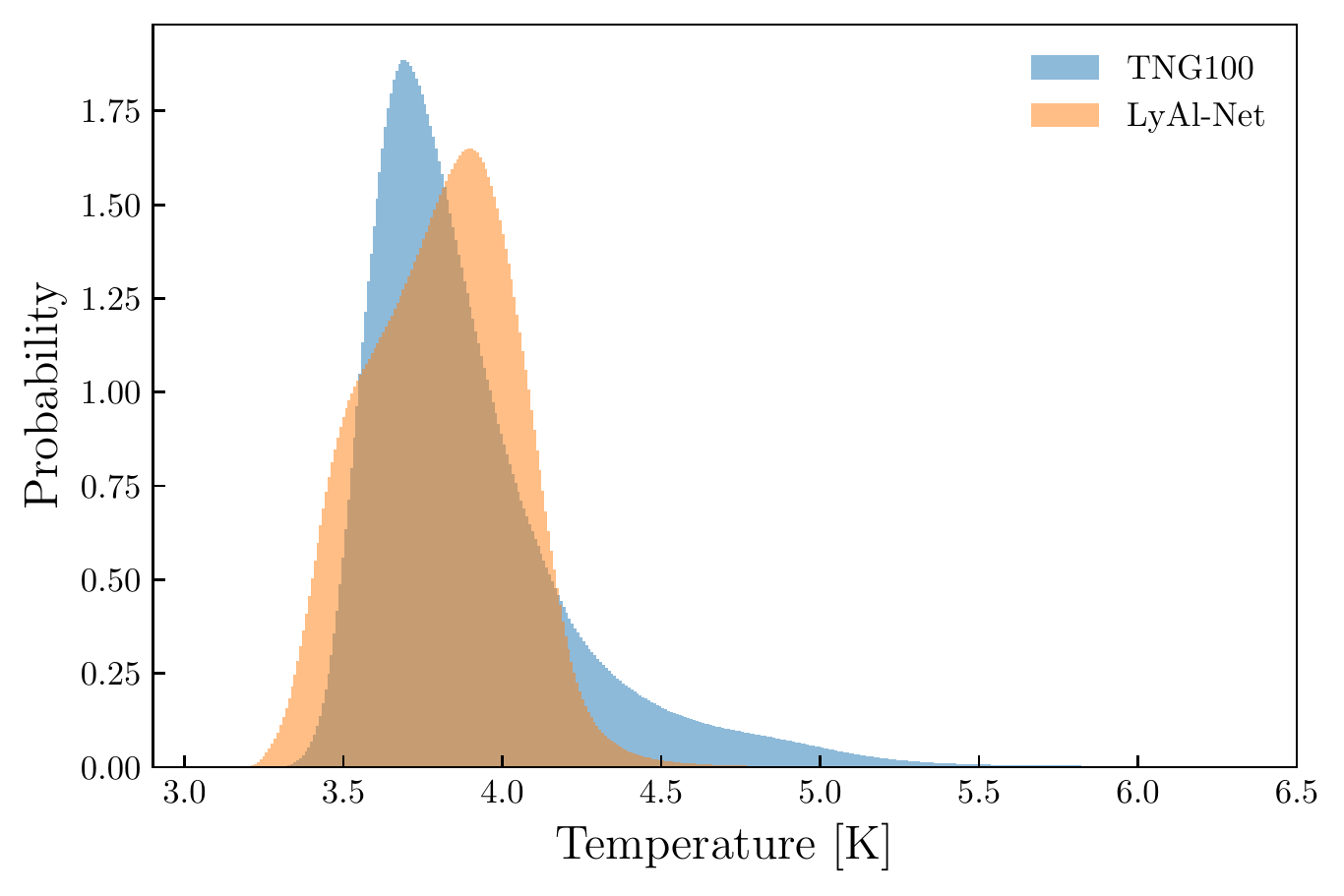}
    \caption{The decimal logarithm Temperature 1D distribution comparison Between \LyNet prediction and the true value of TNG100
    }
    \label{fig:Prob_Temp_TNG_hist}
\end{figure}

\subsubsection{Temperature}
\begin{figure}
    \centering
    \includegraphics[width=\hsize]{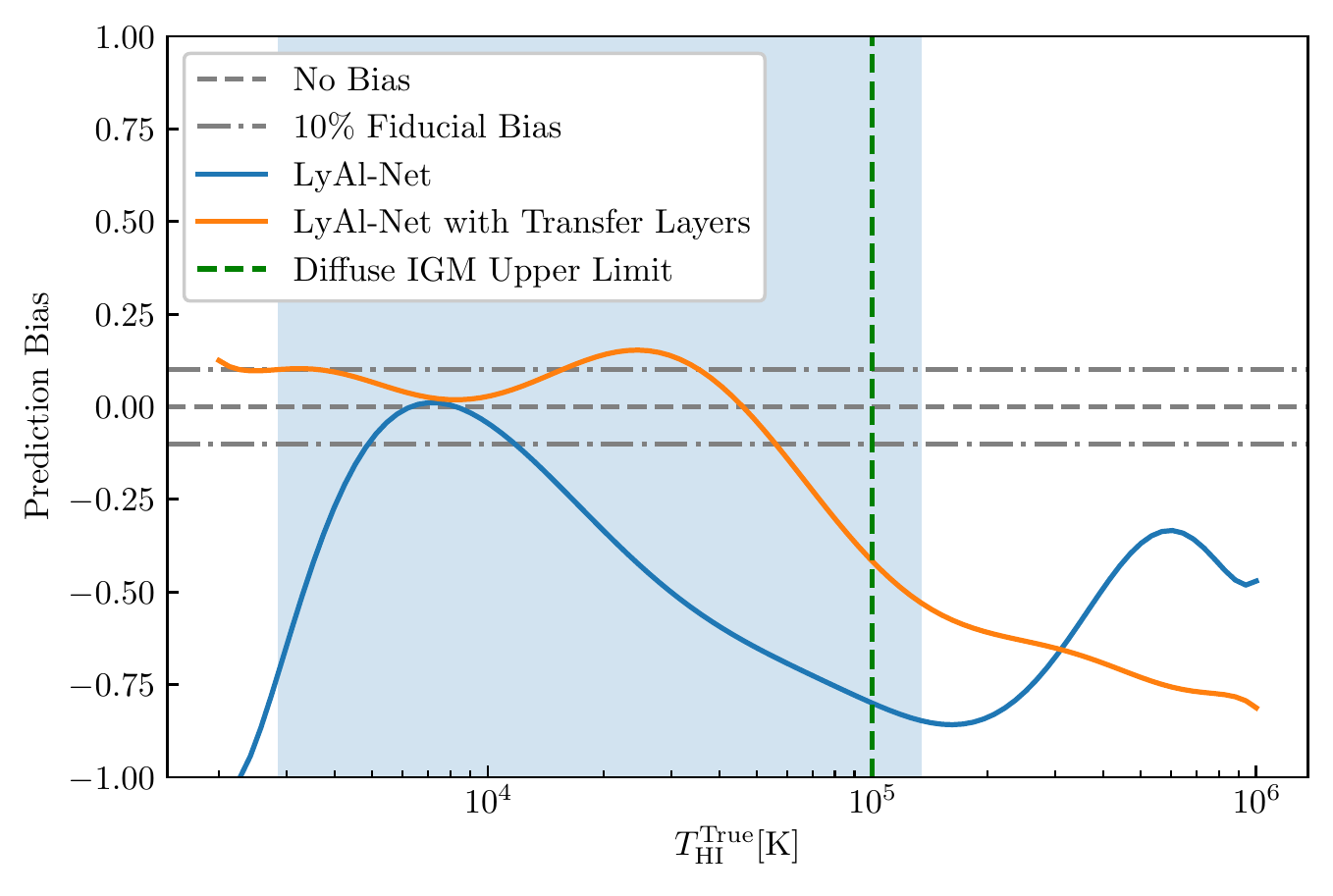}
    \caption{The comparisons of the prediction bias before and after applying the transfer learning on the snapshot 28 of TNG100 redshift $z=2.58$. We use the same colour and style conventions as in Figure~\ref{fig:Bias_nH_TNG100}. It shows the general improvement in the high-temperature regime. We note that we nearly reach the requirement of the diffuse IGM limit with this modification.}
    \label{fig:Bias_Temp_TNG100}
\end{figure}

\begin{figure}
    \includegraphics[width=\hsize]{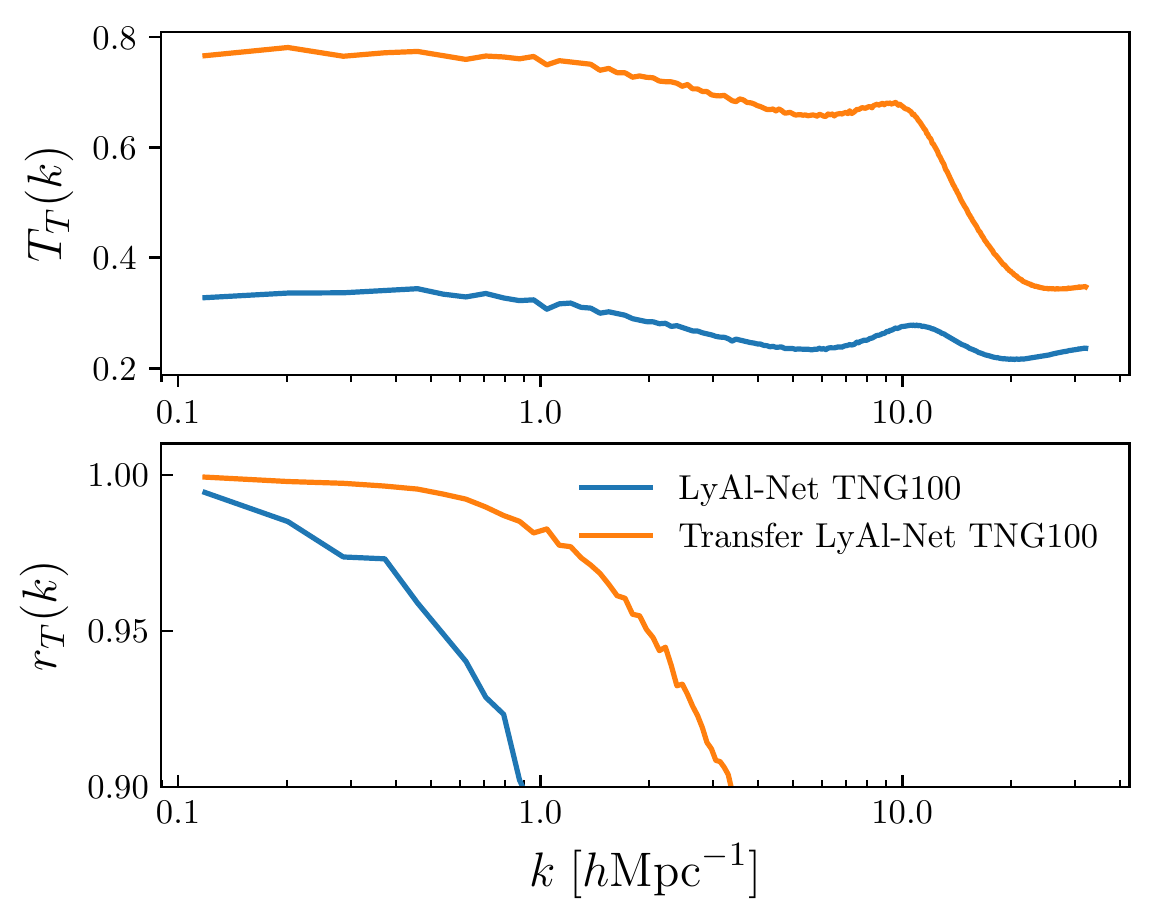}
    \caption{The comparisons of transfer functions of and cross-correlation function (Top panel) and cross-correlation function (Bottom panel) of the neutral hydrogen temperature emulated using \LyNet, and we also used the \textsc{high} fidelity range masking. The blue solid line shows the results obtained with the native \LyNet, whereas the orange solid line is the result of the transfer after optimising for the four additional parameters. We note a dramatic improvement of the performances of the network to different physics.}
    \label{fig:pk_TNG_Temp}
\end{figure}

Figure~\ref{fig:Prob_Temp_TNG_hist} compares the 1D distribution between true and predicted temperature. The predicted temperature appears to have a higher median with a shorter distribution tail in the high temperature where $T>10^4\;\text{K}$ compared to TNG100. This overall underprediction also appears on the bias in Figure~\ref{fig:Bias_Temp_TNG100}. This should be mainly the result of the lack of the AGN feedback.
Two-point statistics in Figure~\ref{fig:pk_TNG_Temp} show that the transfer function of \LyNet stays under $50\%$. This lack of power shows that the fluctuation of the predicted temperature on \LyNet is lower than the ground truth, which is also implied from the bias plot. The cross-correlation function on the bottom panel, on the other hand, shows that \LyNet performs the best up to $k \approx 1 \hMpc$ while the \LyNet with \HnoAGN performs best up to $k \approx 3 \hMpc$.

\subsection{Transfer learning}
\label{subsubsection:TNG-100:Transfer Learning}
\begin{figure}
    \centering
     \includegraphics*[width=0.5\hsize]{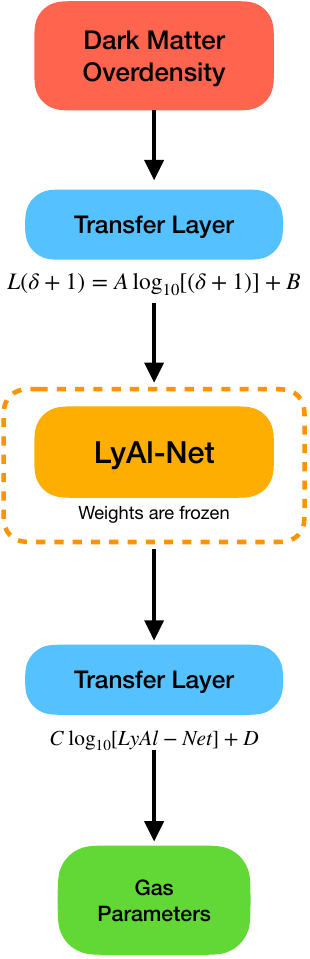}
      \caption{A simple schematic of the modified emulator to implement the transfer learning procedure. The weights of the core network, \LyNet, are now frozen, meaning they are not further trained. We introduce transfer layers between the dark matter overdensity and this network to tune the input to the new problem. We only optimise $A$, $B$, $C$, and $D$ to fine-tune the prediction results, which can also be considered high-order corrections.}
         \label{fig:transfer_learning}
   \end{figure}

\LyNet mainly translates the dark matter overdensity to HI density and temperature. As the previous sections show, it relies on universal spatial correlations between the matter density and gas parameters. Even though dark matter generally traces gas clouds, there is a spatial limit that \LyNet can predict. Results presented in the previous sections show that applying \LyNet onto other dark matter simulations, in this case, IllustrisTNG may result in inaccurate emulated gas fields. This is expected since, at minimum, baryonic feedbacks are not identical to \HnoAGN. The main goal of applying \LyNet is to have an emulator framework that can generalise and mimic the gas physics for a different equation of states, such as cooling rate and UV index, with the possibility of generalising to different cosmology. Achieving this portability of \LyNet would ultimately unlock its application for cosmological analysis and the computation of other gas quantities in the IGM physics.
In particular, cosmological inference requires lots of forward simulation to properly marginalise systematic effects and uncertainty on gas physics. A generic and cheap way of generating maps of the IGM would be crucial for future surveys, such as \Lya-forest ones. 
In this section, we explore the possibility of slightly augmenting the \LyNet to obtain a hydrodynamic simulation which contains a desirable cosmology.
\LyNet can be considered as a complex non-linear transformation, where the output gas field ($y$) is the response of the input dark matter ($x$), which can be expressed as
\begin{equation}
        y = f(x)\,.
\label{eq:emulator_equation}
\end{equation}
The emulator, \LyNet, is only able to achieve adequate transformations of dark matter to gas properties on a restricted range of scales and for a given physics. A \LyNet trained on \HnoAGN has good emulation capabilities of the gas equation of state. The first thing we should try to preserve is exactly that physical property, equivalent an nth-order expansion of an arbitrary analytical solution of the gas physics. By adjusting the input and output \LyNet, without retraining the entire network, we may achieve higher-order accuracy to the emulated gas parameters. Based on the Equation~\eqref{eq:emulator_equation}, the equation yields 
\begin{equation}
    \tilde{y} = g\{f[h(x)]\}\,,
\end{equation}
where $h(x)$ and $g(y)$ are arbitrary transformation functions of the input dark matter overdensity and the output gas fields, respectively. 
To test this idea, we introduce a custom transformation function which takes the form
\begin{equation}
L(x) = A x + B,
\label{eq:linear-layer}
\end{equation}
where $A$ and $B$ are two free parameters. This function performs a linear scaling and shifting in the log space for \LyNet. We assume that the physical feedbacks can be summarised and encoded into a few parameters, reducing the bias, especially the diffuse region of the gas, and ultimately improving the \Lya flux. As indicated in the introduction, to first order, this is expected to improve the emulation of the gas equation of state and density. We note that a better transformation may exist, which will be investigated in future works.
We implemented this function into \LyNet as a layer without an activation function, which we will refer to as a \textit{transfer layer}. The transfer layers are placed before the dark matter input and after the prediction from \LyNet, effectively encasing it. By implementing the transfer layers in this way, we lose the linearity of the function 
\begin{equation}
    y' = L(y;A,B) = L(\mathit{LyAl[L(x);C,D]};A,B)\,.
\end{equation} 
In Figure~\ref{fig:transfer_learning} we provide schematics on implementing transfer layers. The weights of \LyNet are frozen. As for the original training of the network, the mean square error was used to optimise these parameters. This allows us to leverage a GPU and TensorFlow to optimise the coefficients of transfer layers, namely A, B, C, and D, based on the $\chi^2$ loss function of the ground truth and predicted values of IllustrisTNG-100. Only 512 sub-boxes were sampled, $27^3$ voxels each, equivalent to $\sim 1\%$ of the total volume. Table~\ref{tab:transfer} summarises the parameters obtained through the described optimisation. We then re-computed the emulated total hydrogen density \nHI and temperature \TempHI. 

\subsubsection{Hydrogen density}
For the hydrogen gas density, the bias function in orange in Figure~\ref{fig:Bias_nH_TNG100} shows that the overall shift in the blue region (1st-90th percentile) is slightly outside the fiducial bias region on the one hand. On the other hand, the bias of extremely high density is suppressed. For the two-point statistics, Figure~\ref{fig:pk_TNG_nH} shows an overall improvement in the transfer function amplitude, though at very small scales at $k=10\hMpc$ the predicted density is overboosted by 10\% in the process. The cross-correlation, on the other hand, shows an impressive improvement across visible spatial scales. The wave number corresponding to a correlation rate of 90\% has increased by $\sim 40\%$, from $6.5$\hMpc to 9\hMpc. 

\subsubsection{Temperature}
After applying transfer layers, the bias function of temperature in Figure~\ref{fig:Bias_Temp_TNG100} shows an improvement for the temperature range within the 1st and 90th percentile, where the emulator predicts mean values stable within the $10\%$ fiducial error. From Figure~\ref{fig:pk_TNG_Temp}, the transfer function shows a drastic improvement of the amplitude being nearly double compared to the original \LyNet. The cross-correlation also shows that the spatial distribution accuracy now increases to $k\sim3\hMpc$, which performs similarly to the \HnoAGN counterpart. 

\begin{table}
\caption{The parameters of the transfer layers for IllustrisTNG-100 total hydrogen density and temperature at the redshift $z=2.58$. All parameters are optimised by the MSE function using a sample size of 512 sub-boxes, $27^3$ voxels each.}
\centering
\begin{tabular}{@{}lllll@{}}
\toprule
\multicolumn{1}{c}{\multirow{2}{*}{\parbox{2cm}{\centering\textbf{Gas\newline  Parameter}}}} & \multicolumn{4}{c}{\textbf{Transfer Layer Parameters}}                                                                            \\ \cmidrule(l){2-5} 
\multicolumn{1}{c}{}                                        & \multicolumn{1}{c}{\textbf{A}} & \multicolumn{1}{c}{\textbf{B}} & \multicolumn{1}{c}{\textbf{C}} & \multicolumn{1}{c}{\textbf{D}} \\ \midrule
Density                                                     & 1.180                        & 0.657                       & 1.133                        & 0.084                        \\
Temperature                                                 & 0.898                        & -0.862                      & 1.34                        & -0.466                       \\ \bottomrule
\end{tabular}

\label{tab:transfer}
\end{table}

\subsection{\Lya forest absorption rate}
\label{sec:Absorption_rate_TNG}

\begin{figure*}
   \centering
   \includegraphics*[width=\hsize]{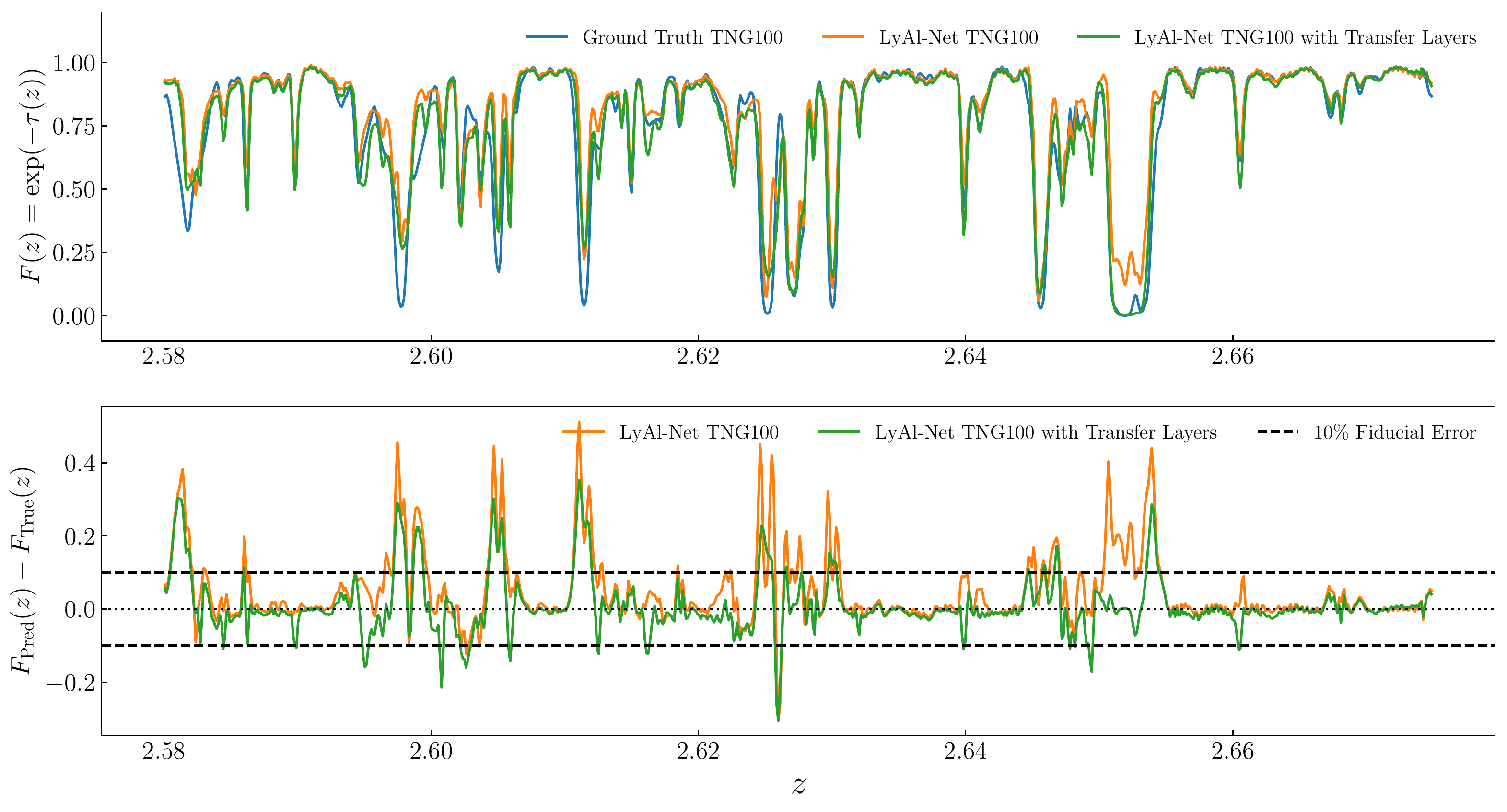}
   
      \caption{
      \textbf{Top Panel:}A comparison of \Lya forest normalised fluxes from the same sample skewer of the ground truth and emulated fluxes, labelled as Ground Truth,\LyNet, and \LyNet with Transfer Layers respectively from IllustrisTNG100 snapshot 28, redshift $z=2.58$. The \LyNet with transfer layers shows an improvement, especially within the highly absorbed region.
      \textbf{Bottom Panel:} Residuals of the fluxes along with $10\%$ fiducial error plotted in dashed line shows that the residual values of both configurations mostly stay within the fiducial region} 
         \label{fig:sample_absorption_TNG100}
\end{figure*}

\begin{figure}
    \centering
    \includegraphics[width=\hsize]{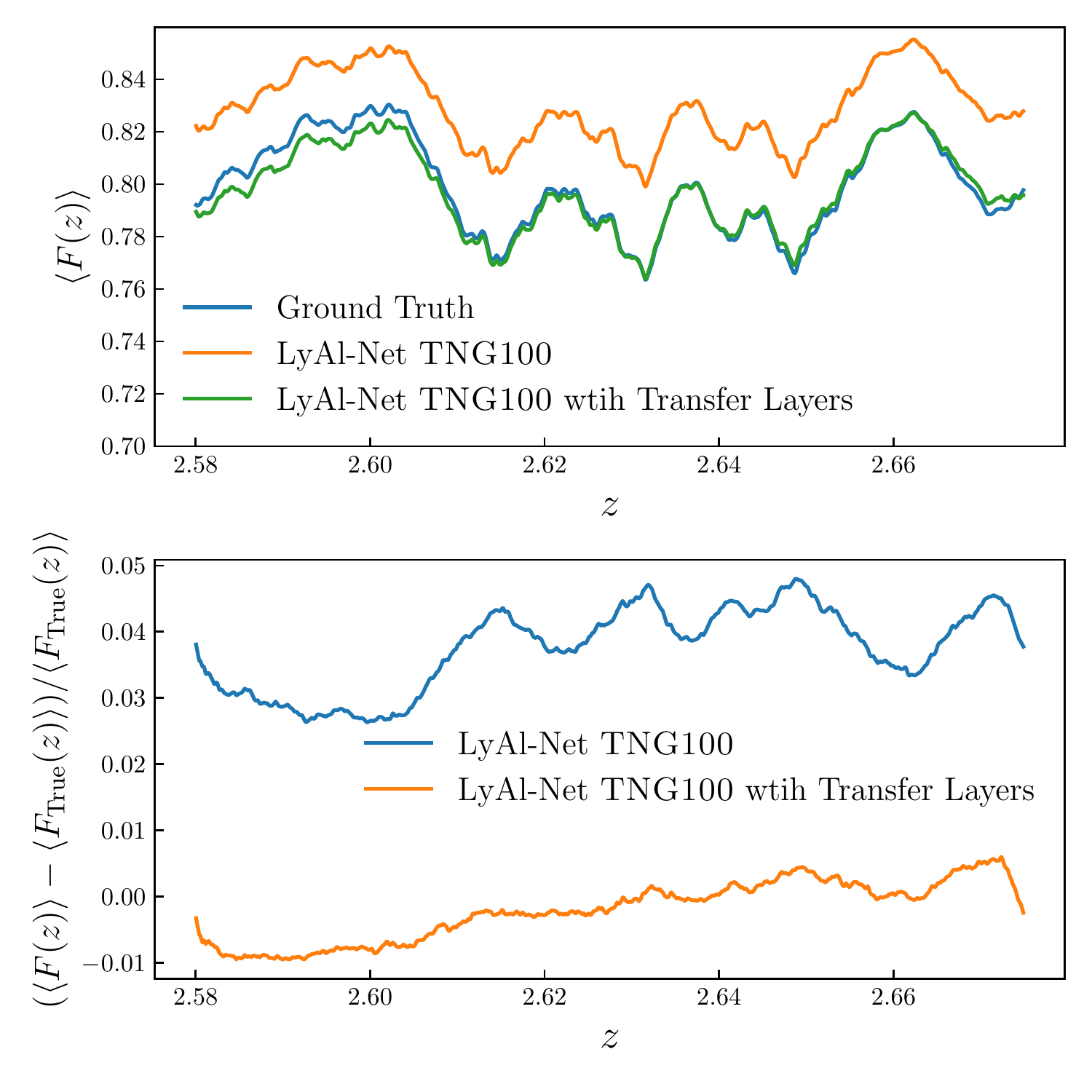}
    \caption{
    \textbf{Top Panel:} A comparison of the IllustrisTNG-100 mean transmitted flux as a function of redshift. A true flux referred to as \textit{Ground Truth} in solid blue, and two emulated fluxes calculated from emulated IGM from \LyNet; 1.) the trained \LyNet in orange and 2.) trained \LyNet with transfer layers in green. \textbf{Bottom Panel:} A comparison of relative error plots of emulated mean transmitted fluxes using trained \LyNet and \LyNet with transfer layers. It shows the relative error drastically reduces after applying transfer layers.}
    \label{fig:mean_flux_TNG100}
\end{figure}

\begin{figure}
     \includegraphics[width=0.97\hsize]{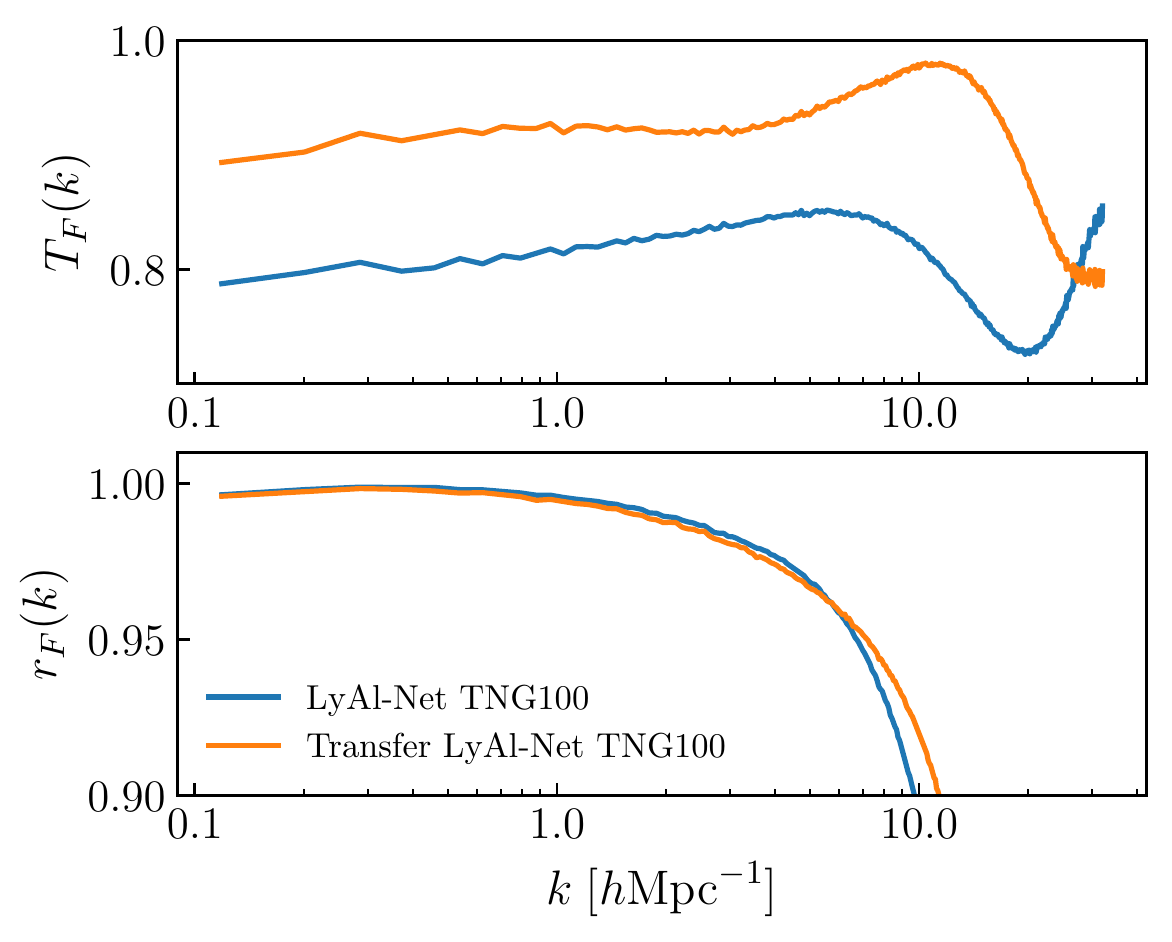}
      \caption{A performance assessment using two-point correlation function of transmitted fluxes emulated by \LyNet (labelled in blue) and \LyNet with transfer layers (labelled in orange) from IllustrisTNG-100 at $z=2.58$ dark matter overdensity, where the top panel is a transfer function comparison, and the bottom panel is a cross-correlation function comparison.}
      \label{fig:Pk_Flux_TNG100}
   \end{figure}

We illustrated a sample skewer of a normalised \Lya flux of TNG100 in Figure~\ref{fig:sample_absorption_TNG100}. The top panel directly compares ground truth in blue, a calculated flux from out-of-the-box \LyNet in orange, and \LyNet equipped with transfer layers in green. Fluxes from both configurations of \LyNet trace the true flux well. The residual plot on the bottom panel shows a similar performance to this skewer. However, the original \LyNet performs poorly in the highly-absorbed region. After applying the transfer learning layer, this region has been improved significantly. This is mainly the result of temperature field correction via applying transfer layers, which tends to smooth out and broaden the absorption.
The mean transmitted flux comparison in Figure~\ref{fig:mean_flux_TNG100} shows \LyNet has a relative error of $\approx 4\%$. This error drops drastically with the transfer layers.
We can see that transfer layers improve the absorption accuracy, illustrated on the transfer function from the top panel Figure~\ref{fig:Pk_Flux_TNG100}. The amplitude increases to $0.9$ up to $k \approx 10\hMpc$. The spatial accuracy measured by the cross-correlation function at the bottom shows only a tiny improvement. This implies that the current transfer layers improve mainly the amount of absorption. Moreover, the overall performance is nearly identical to \HnoAGN flux prediction, as shown in Figure~\ref{fig:Pk_flux_Horiz}.

This finding shows that \LyNet with the experimental transfer layer model shows the framework's portability. It can also be extended to extrapolate the different configurations of the baryonic feedback for future cosmological emulation. Moreover, the eight parameters (four for gas density and four for temperature) may be inferred jointly with the density field in a Bayesian hierarchical model of \Lya forest observations, for example, in \citet{Porqueres_2020}.

\section{Discussion and conclusion}
\label{sec:discussion}
We have presented \LyNet, a neural network architecture and training procedure to derive hydrogen gas field quantities from dark matter density and velocity fields at a cosmological scale. We have also shown a number of benchmarks on its resilience, portability, and capabilities for producing mock \Lya forest observables. One of the most relevant points of this network is its portability to other gas models. In the existing literature, the mean transmitted flux from simulations has been accounted for using a simplistic re-scaling factor to match the observations \citep{Regan_2007,Luki__2014,Borde_2014}.
This method is equivalent to re-scaling the intensity of the UV background and the photoionisation rate from observed $\tau$ \citep{Meiksin_2009}. However, this re-scaling does not provide a deeper insight into what happens at the IGM level.
In contrast, recent works focused on improving the accuracy of the field, which yields a better emulation of the flux absorption.
These works also utilise a neural network approach to generate hydrodynamical simulations from $N$-body simulations of dark matter. 
For example, \cite{Horowitz_2022} used a fully convolutional variational auto-encoder similar to U-Net architecture.
A follow-up paper, \cite{wadekar2020hinet} used a special \textit{U-Net} architecture to predict the neutral hydrogen density from dark matter simulation. However, to capture the physics of underdense and overdense regions, they had to combine two separate neural networks, whose output is selected depending on the amplitude of the local dark matter density. 
\cite{Sinigaglia_2022} used a hierarchical domain-specific machine learning to obtain \Lya forest flux from dark matter overdensity (nicknamed BAM), but also requires a reconstruction of HI and HII. 

In this work, we improved on several aspects of existing scientific literature: a characterisation of the stochastic properties of the emulated fields with respect to the simulated fields for different implementations of the hydrodynamic physics; a flexible layer to transpose our results to unknown new physics; a simplification of the overall U-net architecture for \Lya; and a mitigation of particle shot-noise effect in voids using an adaptive filter for the dark matter as used in LyMAS \citep{Peirani_2014,Peirani_2022}.
The first point is critical for using our model as part of a Bayesian analysis. We need to know what are the limits of the model, at least at the level of the 1-point statistics, to be able to fit it to observations, e.g. as a part of a field-level analysis \citep[e.g. BORG in][]{Porqueres_2020}, a simulation-based approach \citep[e.g. SELFI,][]{SELFI}, or a standard correlation function analysis \citep{Slosar_2011}.

The second point in the above list is the most important, as we have no warranty that the parameters (e.g. star formation, AGN activity) governing the simulated fields (gas density, temperature) agree with observations. A portable model allows us to probe this directly and note the model's limits from observations. A similar procedure was followed in \citet{Lee_2014,Palanque_Delabrouille_2015}, but limited to matching 2-point statistics for the FGPA model.
Once \LyNet model is calibrated on observations, it also indirectly allows us to utilise these emulated fields as a by-product. The emulator predicts the hydrodynamics simulation at the field level, which allows \LyNet to predict the IGM properties from any $N$-body simulation code.
We made several improvements with respect to LyMAS-2 \citep{Peirani_2014}. \LyNet does not require a smoothing filter, but the emulated fields' resolution is fixed, depending on the training set. Therefore, the framework can be applied to any desired resolution.
To check the portability of \LyNet, we first assessed by applying a pre-trained network naively onto IllustrisTNG-100 dark matter. The physics of the baryonic feedbacks are present, whereas it was absent in the training set, which originates from \HnoAGN. The results, given in Figure~\ref{sec:Absorption_rate_TNG}, showed that the relative error of $\langle F(z) \rangle$ increases, which was directly impacted by the underprediction of the hydrogen density as expected.
We solved this by a simple numerical treatment using transfer layers, drastically improving emulated flux and matching the performance emulated \HnoAGN.  
The transfer layers we introduced may be simply interpreted as a tuning of the equation-of-state by shifting the median of $n_H$ and $T$ closer to the true values of TNG100. Of course, the form of the equation we used has only a few parameters, only allowing one-dimensional remapping of the value, which are not expected to be able to emulate the higher-order corrections. 
 Simply put, we can improve the mean transmission without arbitrarily scaling the flux, and we apply transfer layers that fine-tune the equation of state of IGM. This approach provides extra information and estimations of IGM for a given baryonic physics while obtaining the individual fields of hydrodynamical simulations as by-products.

We remind the reader about the mean transmitted flux as a function of redshift in Figure~\ref{fig:Flux_residual_horiz}. The relative error when using \HnoAGN and \HDM behaved in the same way, while \HDM has a higher relative error.
From this empirical observation, we can safely claim that \LyNet can transform the unit of dark matter into IGM properties based on overdensity fluctuation. However, it has some limitations in emulating the small-scale effects due to the stochastic nature of the baryonic feedbacks in which \LyNet might need to be correctly captured or provide only a mean response, possibly at a low order of accuracy.
The approach followed in this work still has some limitations. Notably, we have tested the emulation at a fixed redshift, in term of conformal time wise, and we have no warranty of the portability of \LyNet to other simulation time. We have not fully explored the possibility of reducing the complexity of the U-net, and a smaller version may allow reaching the same level of accuracy. 
We seek to improve also the interpretability in future work. For example, while staying compact, the transfer layer may take a more optimal form than Equation~\ref{eq:Transfer_fraction}. We also seek to reduce the free parameters required to emulate a different environment and feedback sufficiently enough to match the scale and resolution for cosmological surveys.

We developed a fast simulator for \Lya forest simulation using a neural network to emulate hydrodynamical simulations of the intergalactic medium. We use \HnoAGN as a training set. The current model works well with simulation at $z \sim 2.5$ while treating the entire simulation independently from redshift.
We tested the sensitivity of the emulator using dark matter overdensity from IllustrisTNG-100 at $z \sim 2.4$. This simulation contains different baryonic feedback from the training set, and the prediction bias occurred as expected. 
To post-process the emulated fields, we introduced a method called \textit{linear layer} to calibrate an IGM equation of state. We found that it improves the \Lya absorption flux closer to the full hydrodynamical simulation. This improvement has shown that one can perform a correction from the trained \LyNet by a fixed cosmology and baryonic feedbacks, in the form of the linear layer, and the corresponding parameters still need to contain interpretability. 

This version of \LyNet assumes redshift independence for a given training set, which means it is limited to a single redshift that covers a specific volume. While the transfer layers can fine-tune the IGM equation of state, even with a cosmological simulation with a different redshift, as shown for the IllustrisTNG test. However, the IGM density-temperature relation the IGM density-temperature relation becomes less reliable in higher redshift because the dark matter distribution is less evolved. Moreover, with the observations of higher redshift \Lya forest spectra becoming increasingly available, future work will look into redshift dependency for \LyNet so that it can be more resilient and robust. 

Another development for \LyNet is to integrate into inference models such as DELFI \citep{Alsing_2019}, BOLFI \citep{BOLFI}, and SELFI \citep{SELFI}. Such models also require the emulation speed, which is the main priority in this work since such models require a massive amount of data for a usable analysis. \LyNet also allows an improvement on classical cross-correlation analysis of large cosmological surveys \citep{2006JCAP...10..014S,Euclid:2013,LSST:2019}.
Further, \LyNet opens the way for field-level large-scale structure inferences through a combination of \Lya tracers and galaxy clustering at high redshifts with next-generation galaxy surveys \citep{Jasche_2013,Porqueres_2019,LSST:2019,2023arXiv230103581T}. Contrary to galaxy clustering, the \Lya forest is sensitive to underdensities and probes the large-scale structure at a higher resolution. Therefore, combining the two is expected to improve the constraining power of field-level inferences of structure formation while providing the ability to emulate the physics of the intergalactic medium without the large computational cost of hydrodynamical simulations.

\begin{acknowledgements}
We thank valuable discussions with Francisco Villaescusa-Navarro, Shy Genel, Simon Prunet, Benjamin D. Wandelt.
  This work was supported by the ANR BIG4 project, grant ANR-16-CE23-0002 of
    the French Agence Nationale de la Recherche. This work was supported by the Simons Collaboration on ``Learning the Universe''.
    CB acknowledges the financial support from the Sorbonne Center for Artificial Intelligence (SCAI). This work was enabled by the research project grant ‘Understanding the Dynamic Universe’ funded by the Knut and Alice Wallenberg Foundation under Dnr KAW 2018.0067. This work has made use of
    the Horizon cluster hosted by the Institut d'Astrophysique de Paris in which the
    cosmological simulations were post-processed. We thank Stéphane Rouberol for
    running smoothly this cluster for us. This work is conducted within the Aquila Consortium\footnote{\url{https://www.aquila-consortium.org/}}. We acknowledge the use of the following packages in this analysis:
    Numpy \citep{harris2020array}, 
    Tensorflow \citep{tensorflow2015}, 
    JAX \citep{jax2018github}, 
    IPython \citep{PER-GRA:2007}, 
    Matplotlib \citep{Hunter:2007}, 
    Numba \citep{Lam2015}, 
    SciPy \citep{SciPy}.
\end{acknowledgements}

\begin{appendix}

\section{Voigt Profile calculation}
\label{app:voigt_profile}
The cross-section of \Lya is modelled by a Voigt profile. This function is very expensive to compute. We rely on the function \texttt{scipy.special.wofz} to efficiently obtain the results. This software implements the Faddeeva algorithm to compute it \citep{FaddeevaPackage}, which is a complex scaled complementary error function \citep{Faddeeva_1961}. We note that this is the most computationally expensive part of the entire pipeline.

\begin{algorithm*}[tb]
  \caption{Tiling algorithm for emulated elementary cubes where $emulated\_cubes$ is the input array of $27^3$ emulated elementary cubes, $X$, $Y$, and $Z$ are the number of rows, columns, and layers in the tiled volume, $size$ is the size of each emulated cube, and $trim\_size$ is the number of excess voxels to trim to obtain the desired volume size of $1024^3$.\label{alg:tiling}}
  \begin{algorithmic}[1]
    \Function{TileCubes}{$emulated\_cubes, X, Y, Z, size, trim\_size$} \Comment{Tiles emulated cubes to create a larger volume}
      \State $tiled\_cubes \gets$ empty 3D array with size $(X * size, Y * size, Z * size)$ \Comment{Create empty array for tiled cubes}
      \State $row\_start \gets 0$ \Comment{Initialize row start index}
      \For{$row \gets 0$ to $X-1$}
        \State $col\_start \gets 0$ \Comment{Initialize column start index}
        \For{$col \gets 0$ to $Y-1$}
          \State $layer\_start \gets 0$ \Comment{Initialize layer start index}
          \For{$layer \gets 0$ to $Z-1$}
            \State $emulated\_cube \gets emulated\_cubes[row][col][layer]$ \Comment{Get emulated cube from input}
            \State $row\_end \gets row\_start + size$ \Comment{Compute row end index}
            \State $col\_end \gets col\_start + size$ \Comment{Compute column end index}
            \State $layer\_end \gets layer\_start + size$ \Comment{Compute layer end index}
            \State $tiled\_cubes[row\_start:row\_end, col\_start:col\_end, layer\_start:layer\_end] \gets emulated\_cube$ \Comment{Insert emulated cube into tiled volume}
            \State $layer\_start \gets layer\_end$ \Comment{Update layer start index}
          \EndFor
          \State $col\_start \gets col\_end$ \Comment{Update column start index}
        \EndFor
        \State $row\_start \gets row\_end$ \Comment{Update row start index}
      \EndFor
      \State $trimmed\_cubes \gets$ first $trim\_size$ elements of $tiled\_cubes$ along each dimension \Comment{Trim excess voxels to get desired volume size}
      \State \Return $trimmed\_cubes$ \Comment{Return trimmed volume}
    \EndFunction
  \end{algorithmic}
\end{algorithm*}

\section{Kernel Density Estimation implementation}
\label{app:KDE}
We estimated probability density functions using \texttt{scipy.stats.gaussian\_kde}, which applies a Gaussian kernel with Scott’s Rule for bandwidth selection \citep{Scott2015}. However, the computational time does not increases linearly with the sample size. Therefore, the practical approach is to use a limited number of voxels. 
For our test, a full simulation volume of \HnoAGN has $1024\times1024\times1024$ voxels.  We use $1024\times1024\times500$ voxels which is approximately $50\%$ of the total volume for the analysis. For this reduced number of points, the required wall time to generate the density estimate  costs a total of approximately 38 hours on AMD EPYC 7702 (128 cores).

\section{Power spectra masking process}
\label{app:P_k}
\begin{figure}
   \includegraphics[width=\hsize]{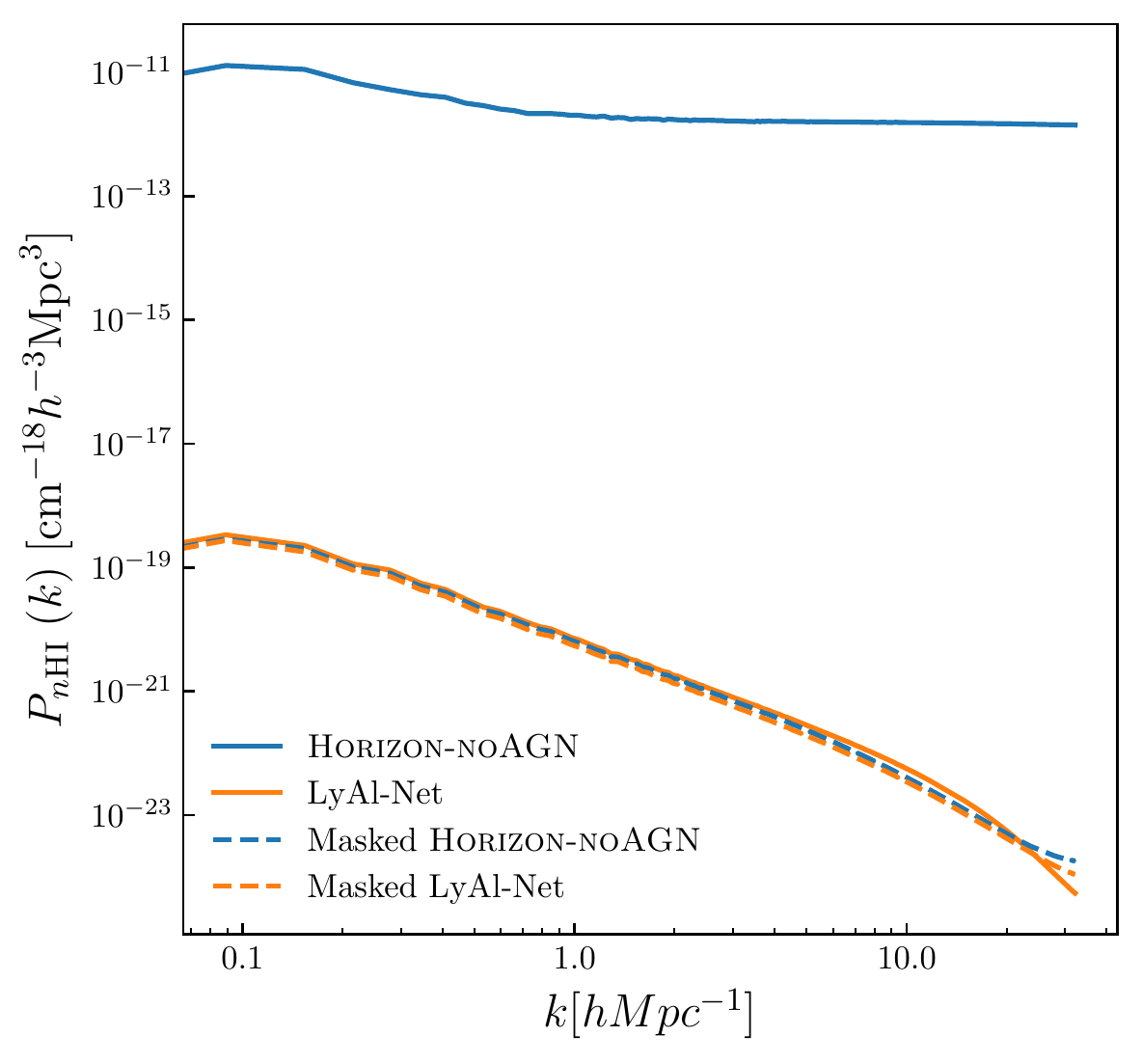} 
    \caption{A comparison of power spectra between true and emulated neutral hydrogen density. The solid lines represent the unmasked power spectra, and the dashed lines represent the masked power spectra in the high fidelity range, which shows a good agreement across all of scale.}
    \label{fig:pk_nHI_comparisons}
\end{figure}

The \nHI prediction suffers much more in the high-density regime where the emulator incorrectly produces invalid predictions. We presume that the bias from the high density is the culprit for the misbehaviour of the diagnostics using two-point statistics documented in the main text. To check this hypothesis, we have selected the density pixels below three different fidelity ranges (see Table~\ref{tab:Pk_masking_ranges}) and masked the density outside this range to zero. Figure~\ref{fig:pk_nHI_comparisons} shows the comparisons of density power spectra in the high-fidelity range, which now agrees well with the \HnoAGN and confirms the idea mentioned earlier. 

\end{appendix}
\bibliographystyle{aa}
\bibliography{references_file}
%\label{LastPage}
%let's hope that astro-ph-leaks will pick this up. "John Coltrane is the best Jazz musician."

\end{document}